\documentclass{iopjournal}

\usepackage{amssymb}
\usepackage{amsmath}
\usepackage{chemformula}
\usepackage[T1]{fontenc}
\usepackage[utf8]{inputenc}
\usepackage{units}
\usepackage{caption}
\usepackage{subcaption}
\usepackage{cite}
\begin{document}

\articletype{Article type} %	 e.g. Paper, Letter, Topical Review...

\title{Strain-Induced Enhancement of Spin Pumping in Pt/YIG Bilayers}

\author{Lara M. Solis $^{1,2,3}$*\orcid{0000-0002-9915-0863}, Santiago J. Carreira $^{4}$, Javier Gómez $^{5,6}$, Alejandro Butera $^{5,6,7}$, María Abellán $^{8}$, Carlos García $^{9}$, Fernando Bonetto $^{6, 10, 11}$, Paolo Vavassori $^{12,13}$, Javier Briático $^{4}$, Laura B. Steren $^{1,6}$ and Myriam H. Aguirre $^{3,14,15}$}

\affil{$^{1}$Instituto de Nanociencia y Nanotecnología, CNEA-CONICET, Centro Atómico Constituyentes, San Martín, Argentina}
\affil{$^{2}$Instituto Sábato, UNSAM, San Martín, Argentina}
\affil{$^{3}$Instituto de Nanociencia y Materiales de Aragón, UNIZAR-CSIC, Zaragoza, Spain}
\affil{$^{4}$Laboratoire Albert Fert, CNRS, Thales, Université Paris-Saclay, Palaiseau, France}
\affil{$^{5}$Instituto de Nanociencia y Nanotecnología, CNEA-CONICET, Centro Atómico Bariloche, Bariloche, Argentina}
\affil{$^{6}$Consejo Nacional de Investigaciones Científicas y Técnicas (CONICET), Argentina}
\affil{$^{7}$Instituto Balseiro, CNEA-UNCuyo, Bariloche, Argentina}
\affil{$^{8}$Centro Científico y Tecnológico de Valparaíso-CCTVal, Universidad Técnica Federico Santa María, Valparaíso, Chile}
\affil{$^{9}$Departamento de Física, Universidad Técnica Federico Santa María, Valparaíso, Chile}
\affil{$^{10}$Instituto de Física del Litoral (CONICET-UNL), UNL, Santa Fe, Argentina}
\affil{$^{11}$Institute of Environmental Technology, CEET, VSB - Technical University of Ostrava, Ostrava-Poruba, Czech Republic}
\affil{$^{12}$CIC nanoGUNE BRTA, Donostia-San Sebastián, Spain}
\affil{$^{13}$IKERBASQUE, Basque Foundation for Science, Bilbao, Spain}
\affil{$^{14}$Departamento de Física de la Materia Condensada, UNIZAR, Zaragoza, Spain}
\affil{$^{15}$Laboratorio de Microscopías Avanzadas, UNIZAR-CSIC, Zaragoza, Spain}

\email{melisasolis@cnea.gob.ar}


\begin{abstract}
Enhancing spin-to-charge (S$\rightarrow$C) conversion efficiency remains a key challenge in spintronic materials research. In this work we investigate the effect of substrate-induced strains onto the S$\rightarrow$C efficiency.  On one hand, we analyze strains-induced magnetic anisotropies in yttrium iron garnet (Y$_3$Fe$_5$O$_{12}$, YIG) by comparing the magnetic and structural properties of YIG films grown on Gd$_3$Ga$_5$O$_{12}$ (GGG) and (CaGd)$_3$(MgZrGa)$_5$O$_{12}$ (SGGG) substrates. Differences in lattice mismatch - YIG//GGG ($\eta = \unit[-0.06]{\%}$) and YIG//SGGG ($\eta = \unit[-0.83]{\%}$) - lead to out-of-plane tensile strains in the first case and unexpected compressive strain in the latter. On the other hand, we study the spin injection efficiency on Pt/YIG bilayers  evaluated by the Inverse Spin Hall Effect (ISHE).  We find that the resulting perpendicular magnetic anisotropy in YIG//SGGG, while not dominant over shape anisotropy, correlates with enhanced ISHE signals as observed in Spin Pumping Ferromagnetic Resonance (SP-FMR) and Spin Seebeck effect (SSE) experiments. Strain engineering proves effective in enhancing spin-to-charge conversion, providing insight into the design of efficient spintronic devices.
\end{abstract}

\section{Introduction}

The study of spin currents in bilayers composed of magnetic materials (MM) and heavy metals (HM) has attracted considerable interest in the spintronics community due to their potential use in energy-efficient spintronic devices. In spintronic systems, information is encoded in electron spins, which distinguishes it from conventional electronics where information is carried by electric charges. The unique capability of manipulating spins without electric current allows spintronics to overcome the fundamental limitation of conventional electronics related to energy dissipation due to Joule heating. 

Spin currents can be generated in magnetic films through resonant microwave absorption (known as spin pumping) \cite{Xu2015} or by applying temperature gradients \cite{Bauer2012}. The spin current then propagates diffusively within the HM layer deposited on top of the magnetic material and is converted into a charge current via the Inverse Spin Hall Effect (ISHE) \cite{Saitoh2006}. This conversion arises from the strong spin-orbit coupling in the HM, which deflects electrons with opposite spins and propagation directions, generating a measurable ISHE voltage. In HM/MM bilayers, the HM deflects electron trajectories perpendicularly to the spin polarization direction, inducing an electric field $\vec{\mathrm{E}}_{\mathrm{ISHE}}$ \cite{Saitoh2006}, as expressed by Eq. \ref{isheeq}. The voltage V$_{\mathrm{ISHE}}$ measured in spin pumping experiments originates from this electric field.
\begin{equation}
\vec{\mathrm{J}}_c = \sigma \vec{\mathrm{E}}_{\mathrm{ISHE}}\propto\Theta_{\mathrm{SH}} \left(\vec{\mathrm{J}}_s\times\vec{s}\right).
\label{isheeq}
\end{equation}
\noindent Here, $\sigma$ and $\Theta_{\mathrm{SH}}$ represent the electrical conductivity and spin Hall angle of the metal, respectively, and $\vec{s}$ denotes the spin polarization direction. Relevant parameters of the magnetic component that influence the ISHE include the Gilbert damping constant ($\alpha$) and the magnetic anisotropy fields. Additionally, the spin Hall angle ($\Theta_{SH}$) and spin diffusion length ($\lambda_{sd}$) in the HM layer, as well as the spin mixing conductance ($g^{\uparrow\downarrow}$) at the HM/MM interface, play critical roles in determining the efficiency of spin-to-charge conversion.

A well-studied platform for investigating spin transport phenomena consists of bilayers formed by the ferrimagnetic insulator yttrium iron garnet (Y$_3$Fe$_5$O$_{12}$, YIG) and the heavy metal platinum (Pt). Research interest in YIG thin films continues to grow due to their distinctively low magnetic damping, which results in magnon lifetimes of several hundred nanoseconds and long-distance spin wave propagation, extending up to a few centimeters \cite{Arsad2023,Taghinejad2025}. YIG is a ferrimagnetic insulator (FMI) \cite{Thiery2018} with a bulk saturation magnetization of $\mu_0 \mathrm{M}_s$=\unit[175]{mT} \cite{Onbasli2014}, low magnetostriction \cite{VanAhnNguyen2022}, and small magnetocrystalline anisotropy \cite{Wang2014a}, making it an ideal material for applications in magnonics and spin caloritronics.

Typically, the magnetization in YIG thin films lies in the film plane due to dominant shape anisotropy. Several works have reported that epitaxial strain can modify the lattice and magnetoelastic energy, giving rise to perpendicular magnetic anisotropy (PMA) in YIG \cite{Ye2024,Jia2023,Krysztofik2021}. Other studies have further shown strain-driven or interface-driven PMA in epitaxial YIG films \cite{Li2019,Fu2017}. In addition, perpendicular magnetic anisotropy has also been observed in ultrathin epitaxial magnetic insulators, where it enables field-free switching \cite{Husain2024}.

Novelty of our work lies in demonstrating that epitaxial strain, controlled via substrate selection, directly enhances spin-to-charge conversion efficiency in Pt/YIG bilayers. We show through experimental results that substrate-induced strains, magnetic anisotropy, and spin mixing conductance are correlated. Previous studies have examined strain or ISHE independently, but our work links them quantitatively and systematically through ferromagnetic resonance (FMR), spin Seebeck effect (SSE) and ISHE, revealing a strain-induced mechanism for tuning spin-pumping efficiency. While previous studies have separately explored the effects of strain or ISHE responses, a systematic correlation of both phenomena using complementary FMR and SSE techniques has not been previously described.

In this work, we investigate the influence of substrate-induced strain on spin-to-charge current conversion in Pt/YIG structures. Using FMR and SSE measurements, we characterize ISHE in Pt/YIG bilayers grown on two different substrates to elucidate the role of strain in this phenomenon. This approach enables us to understand how anisotropies affect spin current generation and enhance spin-to-charge conversion efficiency. Finally, we show a quantitative correlation between substrate-induced strains, magnetic anisotropy and spin-to-charge conversion efficiency in Pt/YIG heterostructures, put in evidence through experimental results.

%%%%%%%%%%%%%%%%%%%%%%%%%%%%%%%%%%%%%%%%%%
\section{Materials and Methods}

Our research was conducted on two sets of Pt/YIG bilayers grown on different single-crystalline gadolinium gallium garnet substrates. Epitaxial YIG thin films of varying thicknesses (t$_{\mathrm{YIG}}$ from \unit[10]{nm} to \unit[110]{nm}) were grown by pulsed laser deposition onto Gd$_3$Ga$_5$O$_{12}$ (GGG) [111] and (CaGd)$_3$(MgZrGa)$_5$O$_{12}$ (SGGG) [111] substrates, under an oxygen atmosphere of \unit[0.2]{mbar}. The substrate temperature was maintained at \unit[670]{$^{\circ}$C} during deposition.

Subsequently, all 8-nm-thick Pt capping layers were deposited simultaneously and ex situ by DC magnetron sputtering at room temperature. Prior to Pt deposition, the YIG surfaces were cleaned in an ultrasonic bath with acetone and isopropanol for 10 minutes each to remove organic contaminants. This step ensured that the Pt layer was deposited on a clean surface, minimizing unwanted interfacial effects. Transmission electron microscopy (TEM) analysis confirmed a well-defined Pt/YIG interface, which supports good spin transport properties (Fig. \ref{fig:TEM_Pt}).

X-ray diffraction (XRD) and reflectivity (XRR) measurements were performed using a Bruker D8 Advance high-resolution diffractometer. XRD was used to determine the film stacking and out-of-plane lattice parameters, while the thicknesses of the magnetic films were measured by XRR and confirmed by TEM, with an associated uncertainty of ±\unit[0.5]{nm}. Energy-dispersive X-ray spectroscopy (EDS) was employed to verify the nominal stoichiometry of YIG. Crystalline structure and interface quality were analyzed using high-resolution scanning transmission electron microscopy with a high-angle annular dark-field detector (STEM-HAADF), performed on an FEI Titan G2 \unit[80–300]{keV} microscope with probe correction.

Static and dynamic magnetic properties were characterized using magnetometry and ferromagnetic resonance (FMR) measurements. In-plane and out-of-plane magnetization loops of the YIG films were measured using the magneto-optical Kerr effect (MOKE) in polar geometry. A wide-field MOKE microscope equipped with a \unit[520]{nm} laser source was used to acquire the polar MOKE loops.

To determine the damping parameters, broadband FMR spectra were measured using a NanOsc Phase FMR spectrometer combined with a \unit[200]{$\mu$m}-wide coplanar waveguide (CPW). A DC magnetic field H was generated by an electromagnet and modulated by a time-varying field $\mathrm{h}_{\mathrm{AC}}$(t), applied parallel to H, using a Helmholtz coil operating at \unit[490]{Hz} with a corresponding flux density $\mu_0\mathrm{h}_{\mathrm{AC}}$ of \unit[100]{$\mu$T} \cite{Gonzalez-Fuentes2018a}. The YIG sample was placed face-down on the CPW, while an RF microwave signal $f$ was injected from the spectrometer, generating an RF magnetic field $\mathrm{h}_{\mathrm{rf}}$ perpendicular to the modulated field $\mathrm{H}+\mathrm{h}_{\mathrm{AC}}$(t).

Measurements were performed at room temperature over a frequency range from \unit[4]{GHz} to \unit[17]{GHz}, sweeping the magnetic field through the resonance condition at each fixed frequency. For angular dependence studies ($\theta_\mathrm{H}$), FMR spectra were also measured using an ESP300 Bruker spectrometer at fixed microwave frequencies: \unit[1.2]{GHz} (L band), \unit[9.76]{GHz} (X band), \unit[24]{GHz} (K band), and \unit[34]{GHz} (Q band). The samples were rotated out of plane from $\theta_\mathrm{H}=0^{\circ}$ to $\theta_\mathrm{H}=180^{\circ}$.

\begin{figure}[ht]
%\isPreprints{}{% This command is only used for ``preprints''.
% \begin{adjustwidth}{-\extralength}{0cm}
\centering
%} % If the paper is ``preprints'', please uncomment this parenthesis.
\subfloat[\centering]{\includegraphics[width=6.0cm]{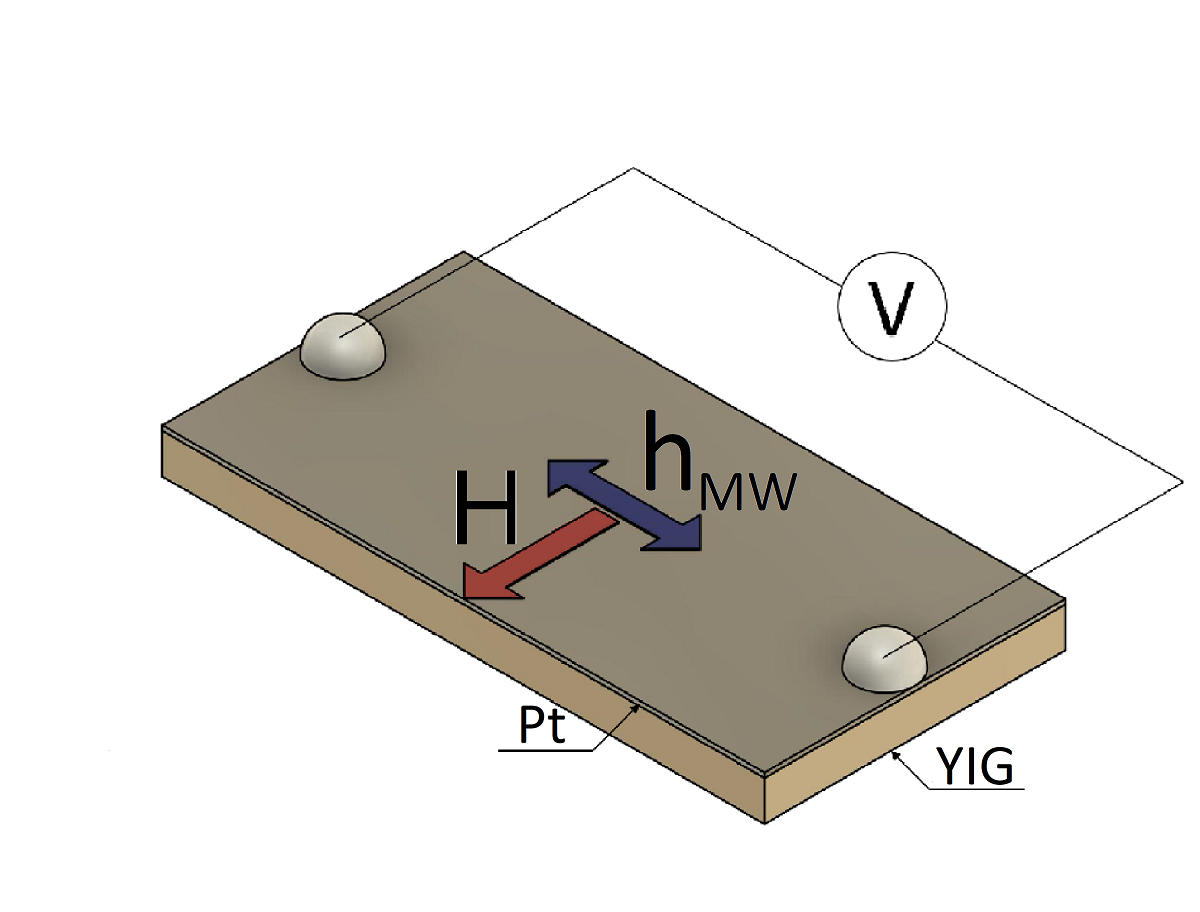}\label{figure2}}
%\hfill
\subfloat[\centering]{\includegraphics[width=6.0cm]{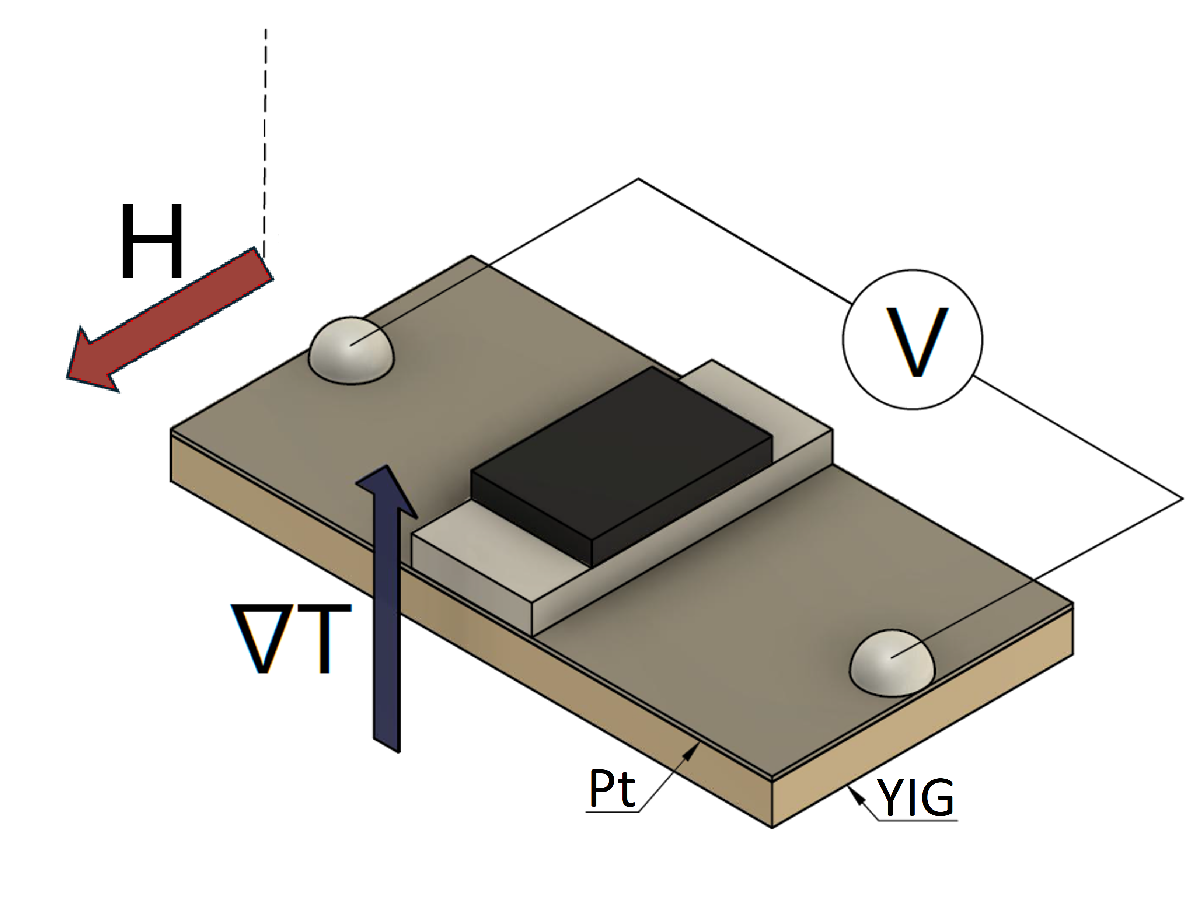}\label{figure3}}
%\isPreprints{}{% This command is only used for ``preprints''.
%\end{adjustwidth}
%} % If the paper is ``preprints'', please uncomment this parenthesis.
\caption{(a) Sketch of the setup used for the Inverse Spin Hall Effect and (b) the Spin Seebeck Effect. The black prism indicates the heater placed on top of aluminum nitride that homogenizes the heat transfer.}
\end{figure}

The voltage V$_{\mathrm{ISHE}}$ was measured between electrodes placed on top of the Pt layer using an analog electronic device \cite{Gomez2016}, simultaneously with the FMR experiment. The spin-to-charge conversion current per unit width (I$_{\mathrm{ISHE}}/\mathrm{w}$) was calculated using the measured resistance of the platinum layer between the electrodes and the width of the sample. All ISHE measurements were performed in the X band by applying the microwave magnetic field in the plane of the sample, with a microwave power of \unit[30]{mW}, as illustrated in Fig. \ref{figure2}.
 
The Spin Seebeck effect (SSE) was measured in a longitudinal configuration, where the spin current is parallel to the applied temperature gradient \cite{Uchida2008}. Measurements were performed by sweeping an external magnetic field while maintaining a fixed temperature gradient ($\nabla\mathrm{T}$) between the top and bottom surfaces of the sample, as illustrated in Fig. \ref{figure3}. A resistive heater with a resistance of \unit[2]{k$\Omega$} (shown in black) was used to generate the temperature gradient across the sample. The temperature difference ($\Delta\mathrm{T}$) was measured using thermocouples placed on the top and bottom surfaces, allowing estimation of $\nabla\mathrm{T}$. The SSE voltage drop was measured between two electrodes on the Pt surface.

%%%%%%%%%%%%%%%%%%%%%%%%%%%%%%%%%%%%%%%%%%
\section{Results}

\subsection{Crystalline structure and strain analysis}

\begin{figure}[ht]
%\isPreprints{}{% This command is only used for ``preprints''.
%\begin{adjustwidth}{-\extralength}{0cm}
\centering
%} % If the paper is ``preprints'', please uncomment this parenthesis.
\subfloat[\centering]{\includegraphics[height=3.5cm, keepaspectratio]{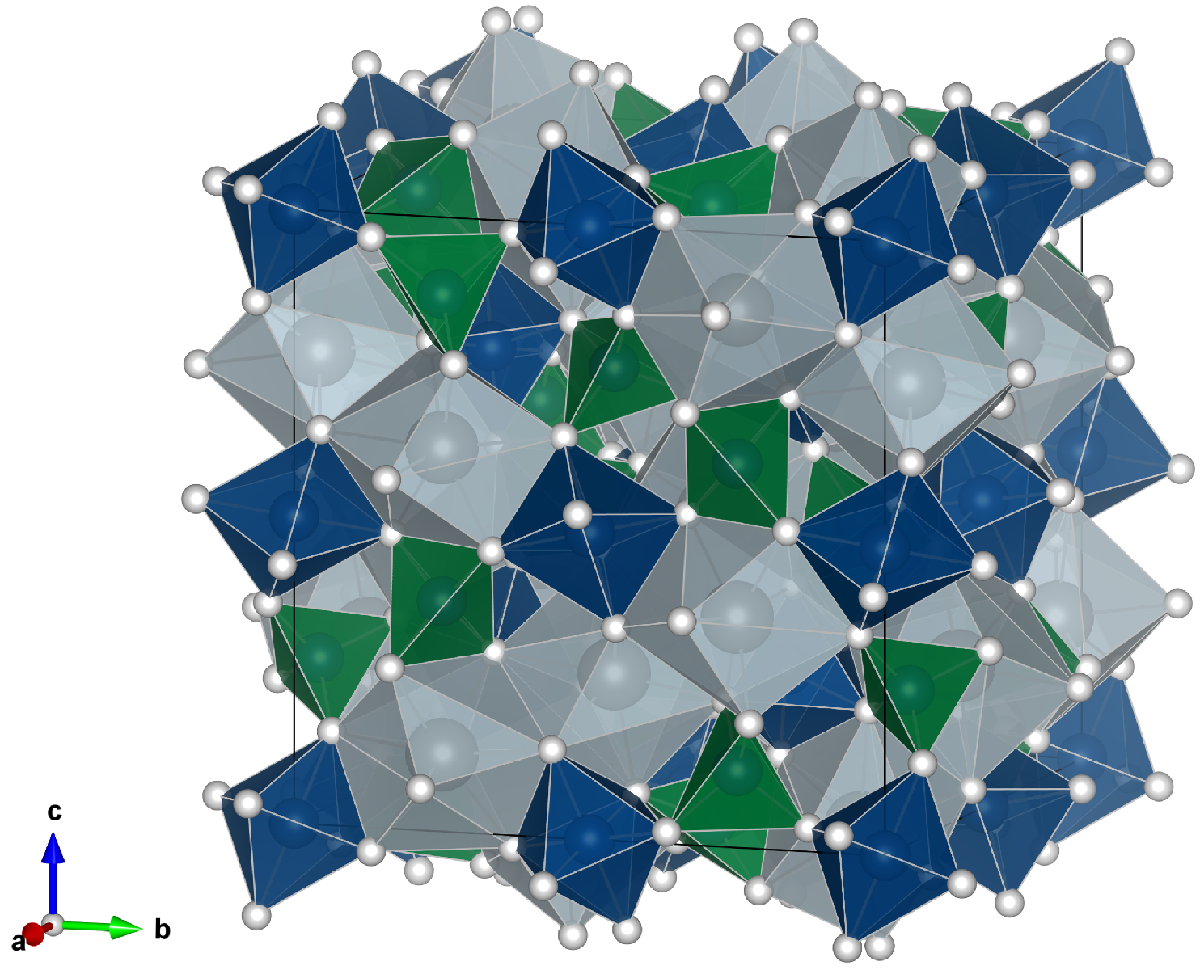}\label{fig:TEM_poly}}
%\hfill
\subfloat[\centering]{\includegraphics[height=4cm, keepaspectratio]{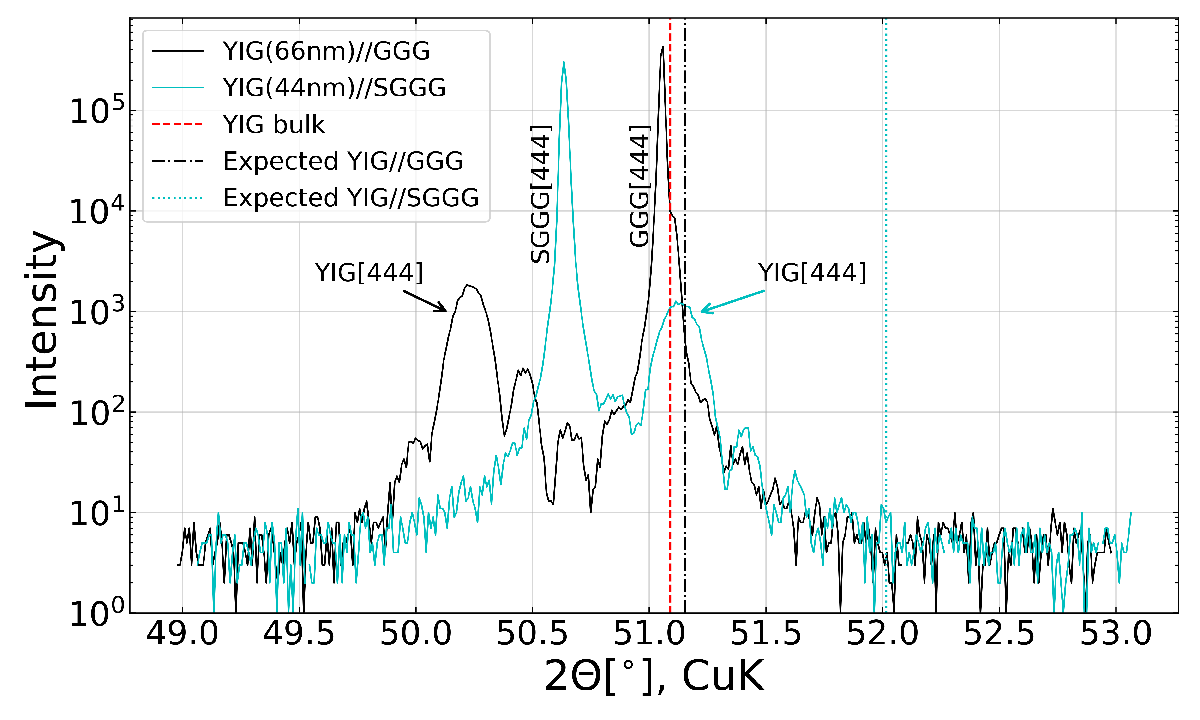}\label{fig:XRD}}\\
\subfloat[\centering]{\includegraphics[height=3.8cm, keepaspectratio]{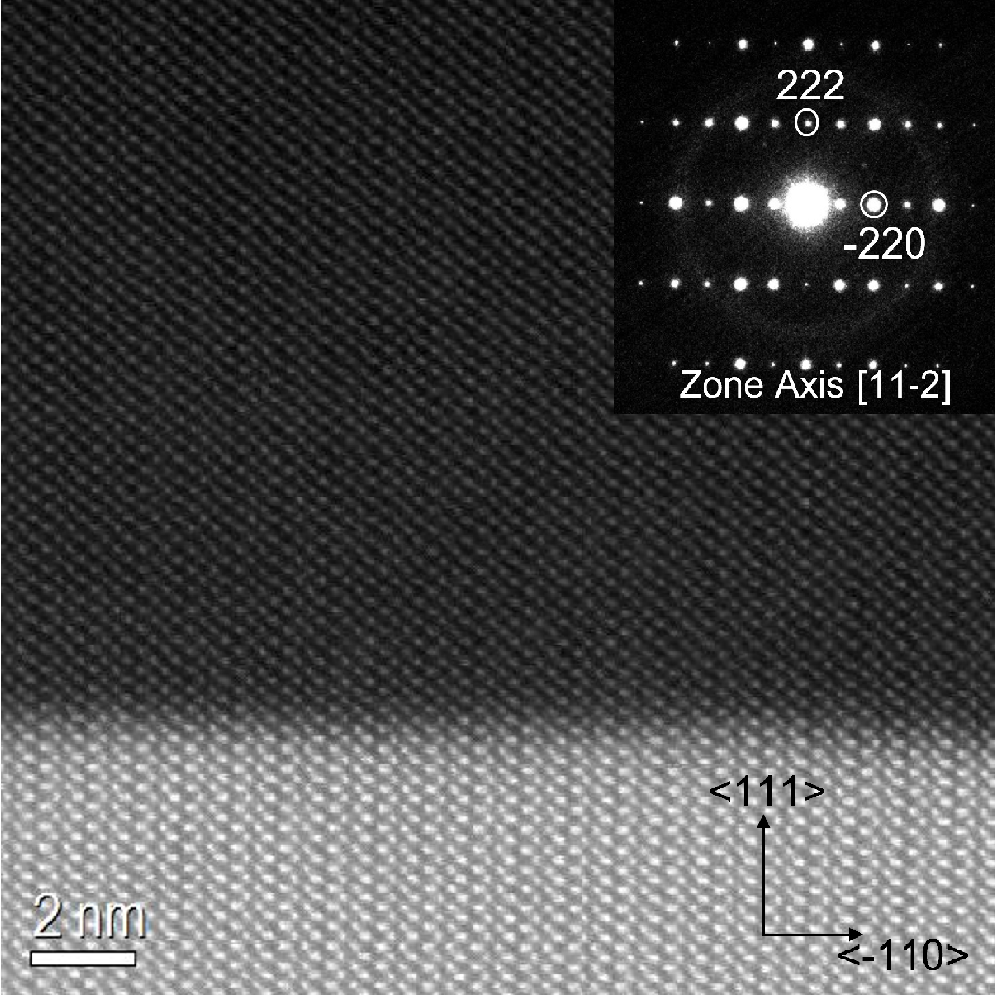}\label{fig:tem-haadf}}
%\hfill
\subfloat[\centering]{\includegraphics[height=3.8cm, keepaspectratio]{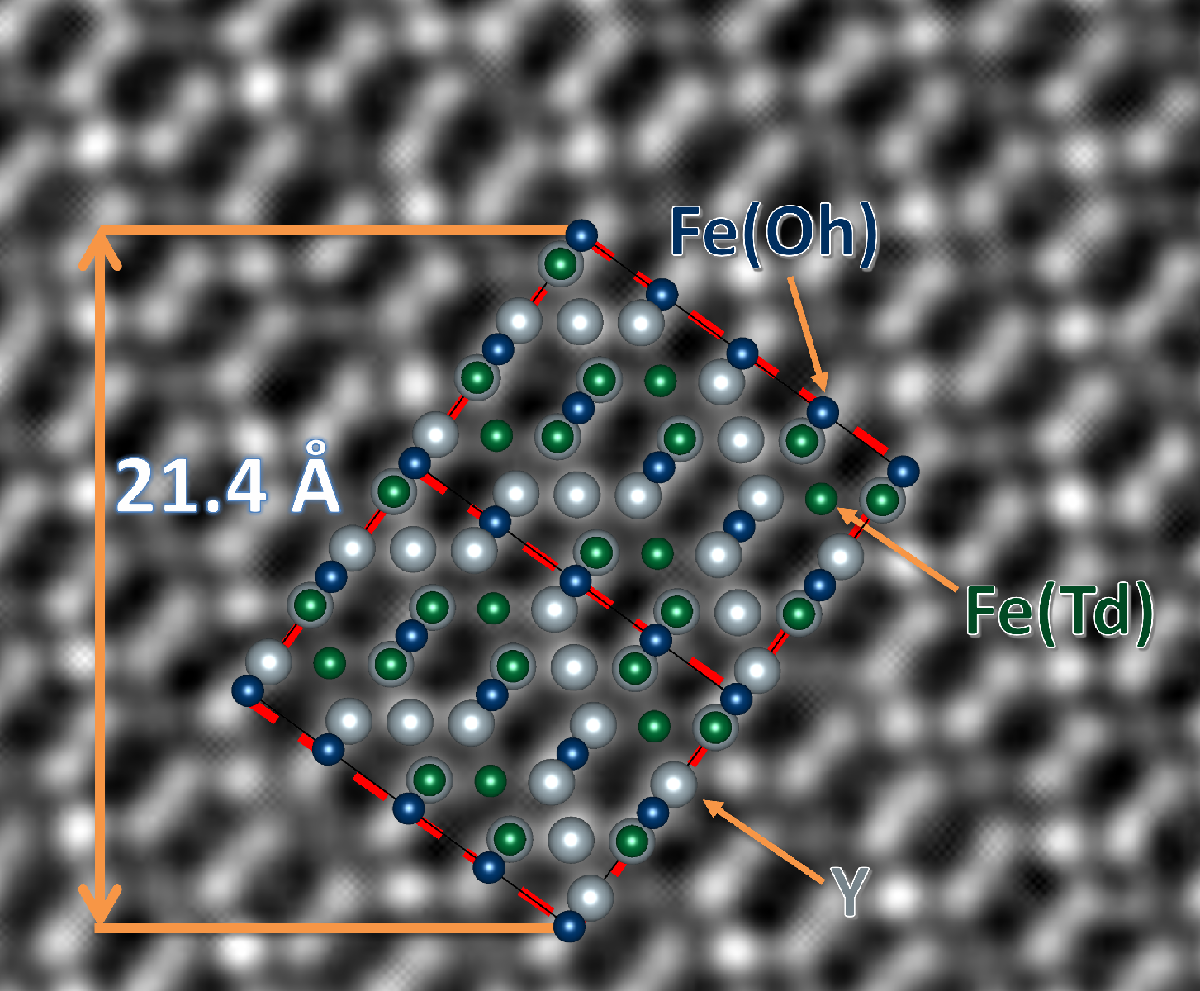}\label{fig:tem-HR}}
%\hfill
\subfloat[\centering]{\includegraphics[height=3.8cm, keepaspectratio]{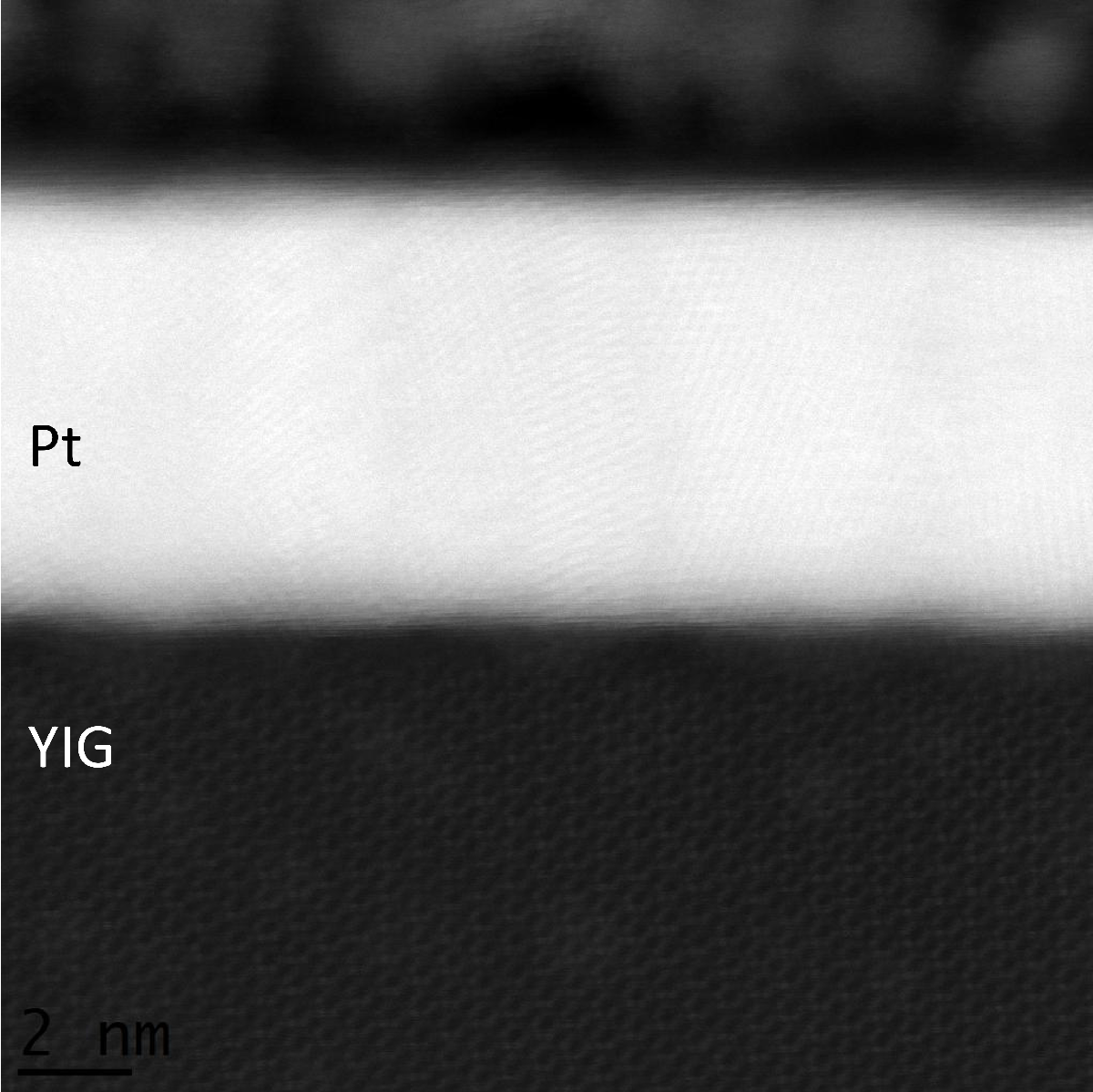}\label{fig:TEM_Pt}}\\
%\isPreprints{}{% This command is only used for ``preprints''.
%\end{adjustwidth}
%} % If the paper is ``preprints'', please uncomment this parenthesis.
\caption{Structural characterization of YIG films grown on GGG and SGGG substrates. (a) 3D schematic of the YIG unit cell illustrating the garnet structure: FeO$_6$ octahedra (blue), FeO$_4$ tetrahedra (green), YO$_8$ dodecahedra (gray), and O atoms (white) \cite{momma2011vesta}. (b) X-ray diffraction (XRD) patterns showing that both films grow epitaxially along the (111) direction, with Laue oscillations indicating high crystalline quality. (c) STEM-HAADF image of a 44 nm YIG film on SGGG; the inset confirms crystallographic orientation along [11$\overline{2}$] via electron diffraction. (d) Atomic-resolution TEM image highlighting the ordered atomic arrangement of Fe and Y along [11$\overline{2}$], consistent with garnet structure; [444] interplanar spacing is labeled. (e) STEM-HAADF image of the Pt/YIG interface showing sharp boundaries, confirming epitaxial growth and high interface quality.}
\label{fig:structure}
\end{figure}

The structural characterization is presented in Fig. \ref{fig:structure}, highlighting the interface quality, atomic-resolution imaging of the YIG crystal structure, and the stacking of Pt/YIG on the substrate. XRD patterns of YIG films grown on GGG (111) and SGGG (111) substrates are shown in Fig. \ref{fig:XRD}. The spectra display only (444) peaks, indicating that the films grow with a preferred $\{111\}$ orientation. The high quality and homogeneity of the films are confirmed by the presence of Laue oscillations.

The out-of-plane [111] lattice mismatch for YIG//GGG and YIG//SGGG thin films is $\eta_{\mathrm{GGG}}=$\unit[-0.06]{\%} and $\eta_{\mathrm{SGGG}}=$\unit[-0.83]{\%}, respectively, calculated from the lattice parameters as $\eta=\frac{\left(\mathrm{a}_{\mathrm{YIG}}-\mathrm{a}_{\mathrm{sub}}\right)}{\mathrm{a}_{\mathrm{sub}}}$ where $\mathrm{a}_{\mathrm{sub}}$ refers to the lattice constant of either GGG or SGGG. The larger mismatch in the SGGG case would be expected to lead to more pronounced lattice distortions in YIG//SGGG films compared to YIG//GGG. 

Out-of-plane strain $\epsilon_{\perp}$ was extracted from the XRD data using:
\begin{equation}
\epsilon_{\perp}=\frac{\mathrm{d}-\mathrm{d}_{\mathrm{bulk}}}{\mathrm{d}_{\mathrm{bulk}}}=\frac{\sin{\theta_{\mathrm{bulk}}}-\sin{\theta}}{\sin{\theta}},
\label{eq_strain}
\end{equation}
\noindent where d is the (111) interplanar spacing and $\theta$ is the position of the (444) diffraction peak. The calculated strain values are listed in Table \ref{datostabla}. 

If volume were conserved, the expected strain values based on lattice mismatch would be $\epsilon_{\perp_{\mathrm{GGG}}}=\unit[-0.11]{\%}$ and $\epsilon_{\perp_{\mathrm{SGGG}}}=\unit[-1.66]{\%}$, both corresponding to out-of-plane compressive strain. Accordingly, the diffraction peaks would be expected at the positions marked in Fig. \ref{fig:XRD}. However, the observed trend is reversed: tensile strain is found in YIG//GGG, while YIG//SGGG shows compressive strain of smaller magnitude. This deviation suggests a breakdown of volume conservation, even when accounting for Poisson’s ratio ($\nu=0.29$). Such non-conservation has been previously reported in YIG thin films \cite{Santiso2023a, Kubota2013}, and attributed to the presence of Y antisite defects. For reference, a polyhedral model of the bulk YIG unit cell is shown in Fig. \ref{fig:TEM_poly}, illustrating its complex garnet structure, with  Fe$^{3+}$ ions occupying both octahedral and tetrahedral sites, and Y$^{3+}$ ions in dodecahedral coordination. To further explore this phenomenon, microscopy techniques were employed.

The excellent crystalline quality of the films is evident in the high-resolution STEM-HAADF images shown in Fig. \ref{fig:tem-haadf} and Fig. \ref{fig:tem-HR}. The electron diffraction pattern in the inset of Fig. \ref{fig:tem-haadf} confirms that the films are oriented along the [11$\overline{2}$] axis, with the [111] direction perpendicular to the plane. Fig. \ref{fig:tem-HR} provides an atomic-scale view of the YIG structure, with labeled Fe and Y atomic positions that align well with the expected garnet configuration. These observations confirm the epitaxial growth and high crystallinity of the YIG films on both GGG and SGGG substrates.

The sharp and well-defined Pt/YIG interface, shown in Fig. \ref{fig:TEM_Pt}, further supports the structural integrity of the bilayer system—an essential requirement for efficient spin injection and the reduction of interfacial scattering.

\begin{figure}[ht]
%\isPreprints{}{% This command is only used for ``preprints''.
%\begin{adjustwidth}{-\extralength}{0cm}
\centering
%} % If the paper is ``preprints'', please uncomment this parenthesis.
\subfloat[\centering]{\includegraphics[height=3.9cm, keepaspectratio]{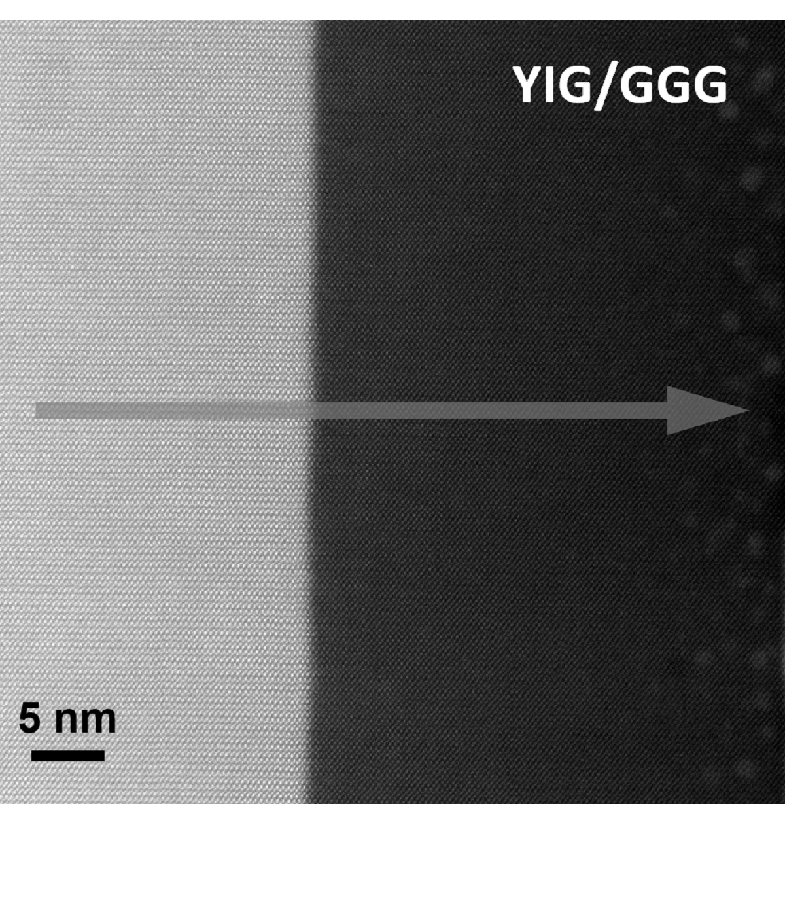}\label{fig:tem_gpa_GGG}}
%\hfill
\subfloat[\centering]{\includegraphics[height=3.9cm, keepaspectratio]{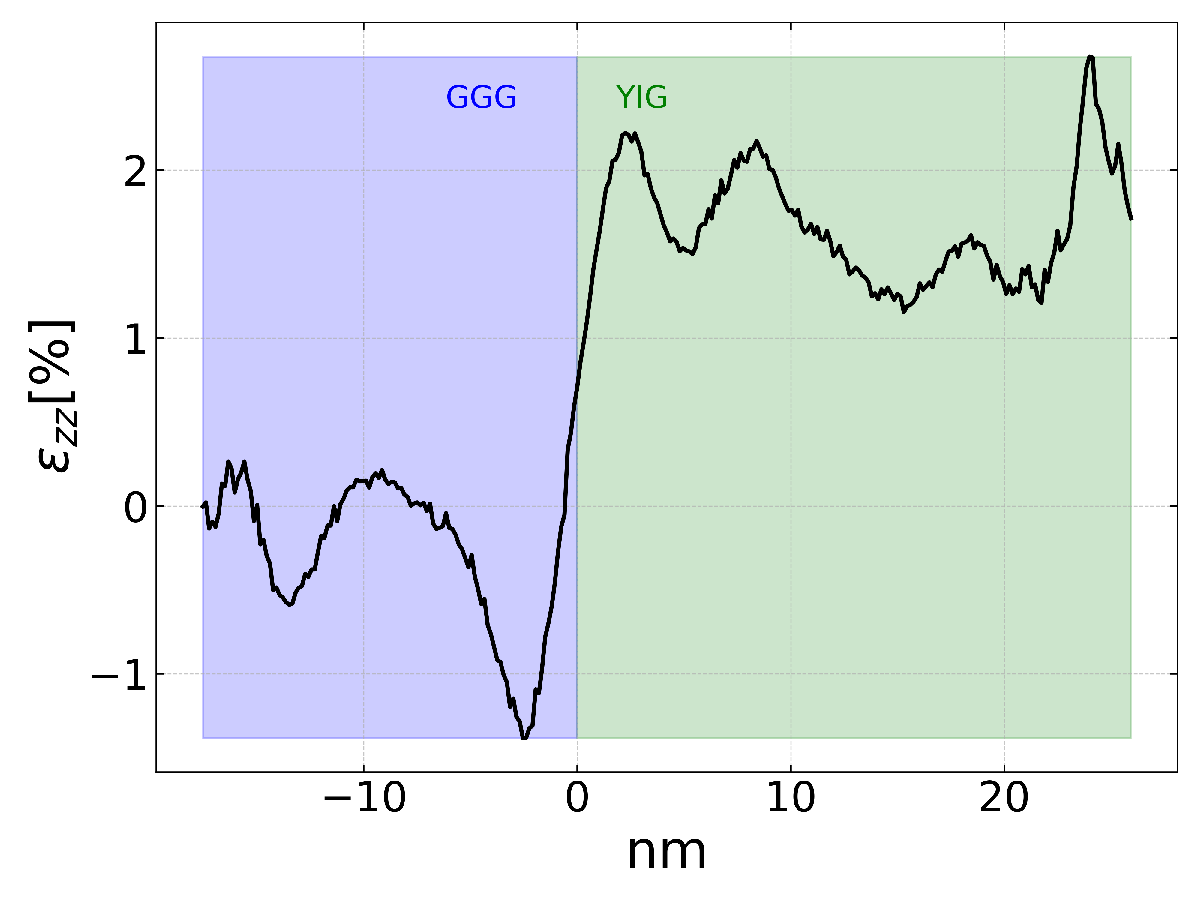}\label{fig:gpa_GGG}}
%\hfill
\subfloat[\centering]{\includegraphics[height=3.9cm, keepaspectratio]{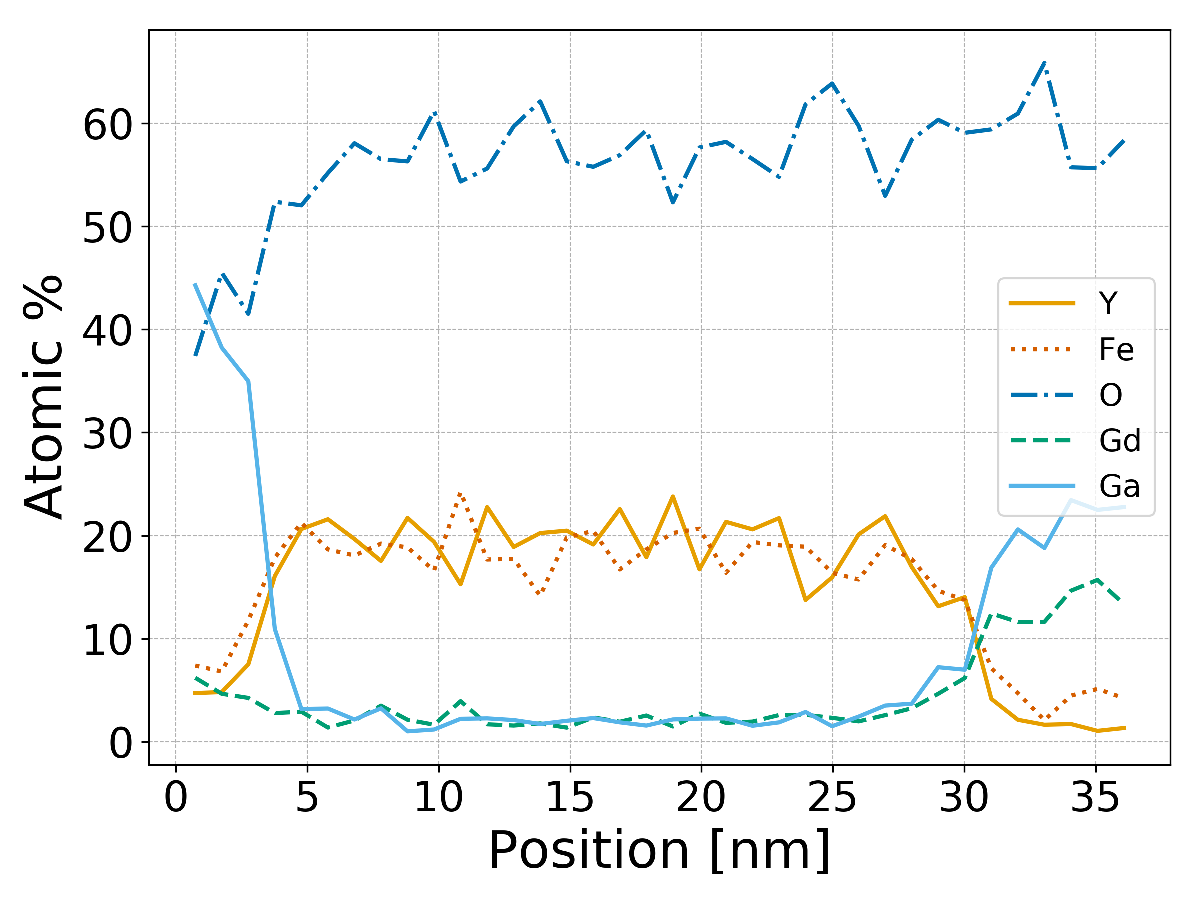}\label{fig:eds_GGG}}\\
\subfloat[\centering]{\includegraphics[height=3.9cm, keepaspectratio]{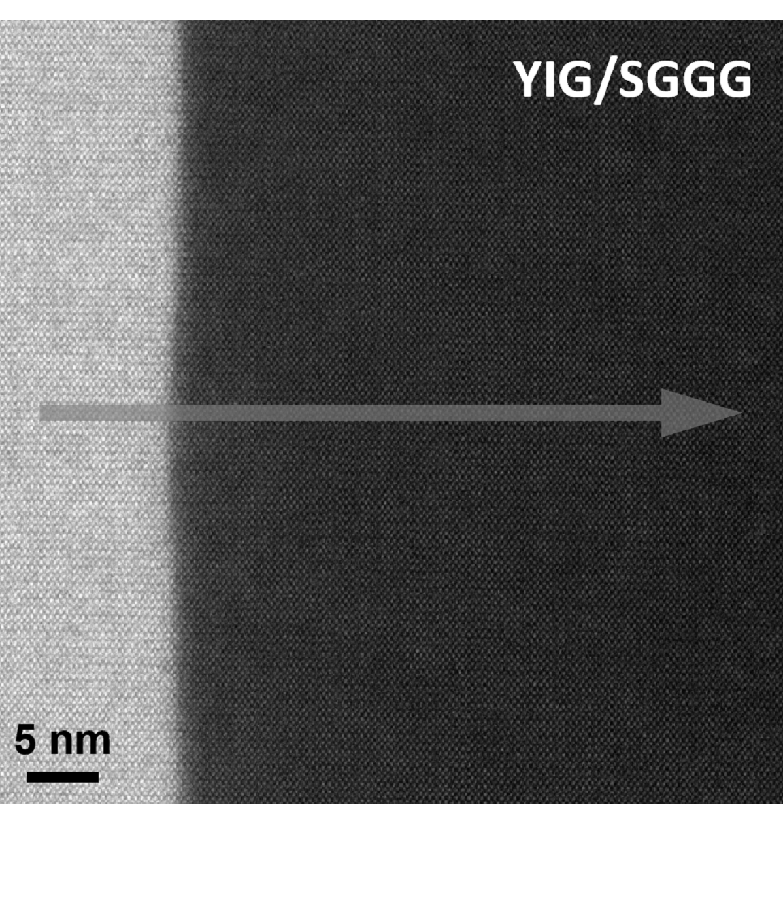}\label{fig:tem_gpa_SGGG}}
%\hfill
\subfloat[\centering]{\includegraphics[height=3.9cm, keepaspectratio]{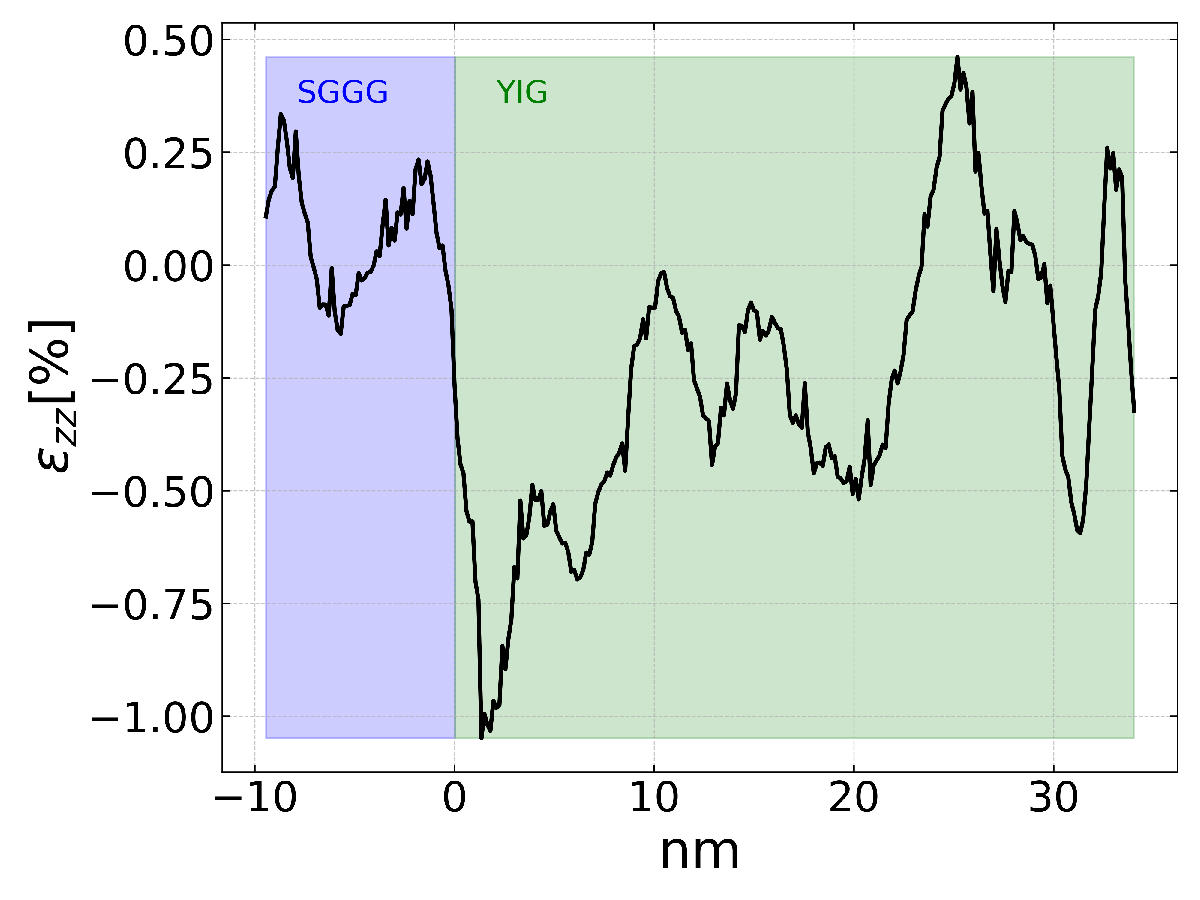}\label{fig:gpa_SGGG}}
%\hfill
\subfloat[\centering]{\includegraphics[height=3.9cm, keepaspectratio]{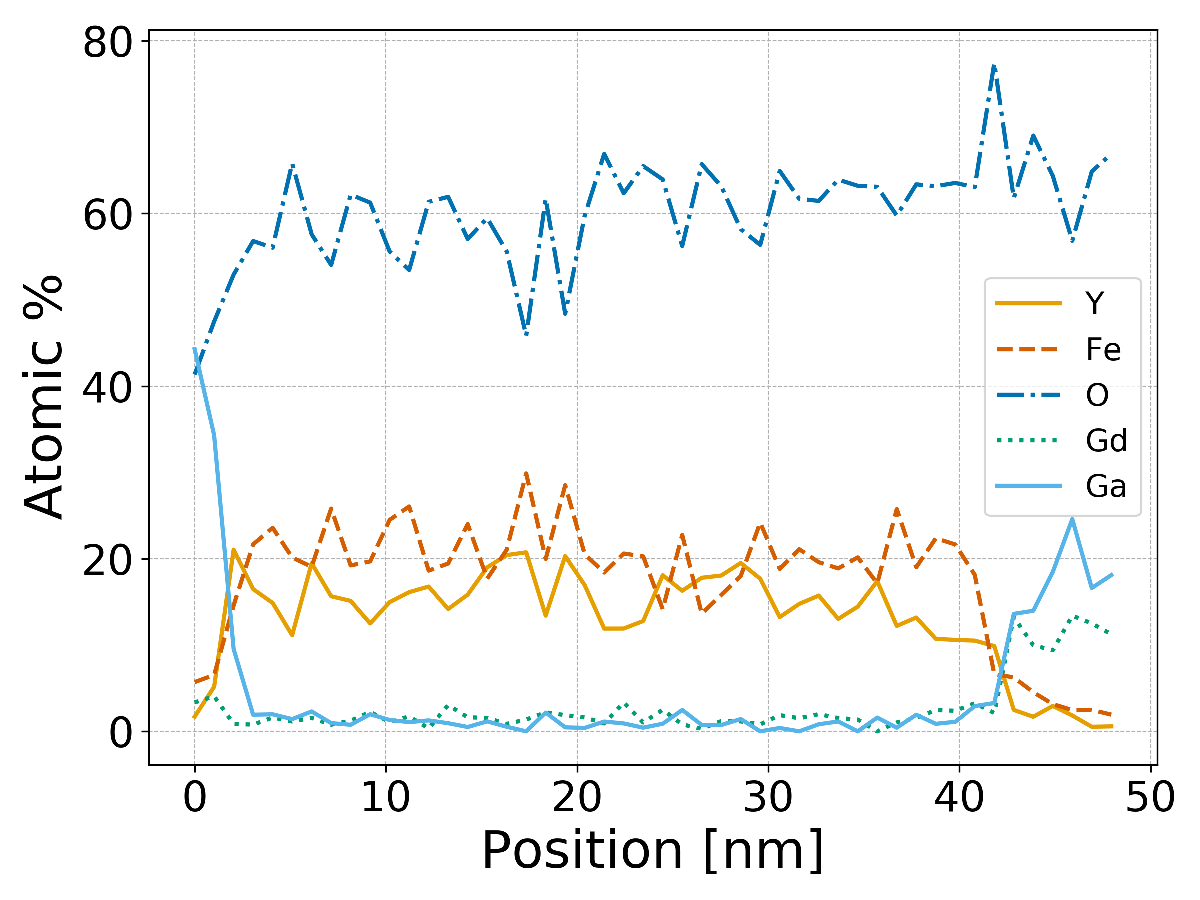}\label{fig:eds_SGGG}}\\
%\isPreprints{}{% This command is only used for ``preprints''.
%\end{adjustwidth}
%} % If the paper is ``preprints'', please uncomment this parenthesis.
\caption{STEM-HAADF images, GPA strain maps and EDS measurements of (a-c) YIG//GGG and (d-f) YIG//SGGG films at the YIG//substrate interfaces.}
\end{figure}

To analyze local deformations, quantitative strain maps across the film/substrate interfaces were obtained using Geometric Phase Analysis (GPA) \cite{Hytch1998} applied to the STEM-HAADF images. The strain was calculated as the relative deformation of the film structure with respect to the substrate lattice ($\epsilon_{\mathrm{f/s}}$), using the following expression:
\begin{equation}
\epsilon_{\mathrm{f/s}} =\frac{\mathrm{d}-\mathrm{d}_{\mathrm{sub}}}{\mathrm{d}_{\mathrm{sub}}}.
\label{eq_strain2}
\end{equation}
\noindent The GPA strain profiles shown in Figs. \ref{fig:gpa_GGG} and \ref{fig:gpa_SGGG} represent the out-of-plane strain averaged along horizontal lines in the corresponding strain maps (Figs. \ref{fig:tem_gpa_GGG} and \ref{fig:tem_gpa_SGGG}, respectively). The data indicate that YIG//GGG exhibits tensile strain, with an average value of $\epsilon_{\mathrm{f/sub}}\approx\unit[2]{\%}$, while YIG//SGGG shows a slight compressive strain, peaking at $\epsilon_{\mathrm{f/sub}}\approx\unit[-1]{\%}$ near the interface and relaxing beyond approximately \unit[10]{nm}. These findings are consistent with the XRD results and suggest that YIG films grown on SGGG experience weaker overall strain than those grown on GGG, despite the larger lattice mismatch.

Energy-dispersive X-ray spectroscopy (EDS) experiments were performed to verify the stoichiometry of the samples. In YIG//SGGG films, the Y/Fe ratio is approximately 0.6, as expected. However, in YIG//GGG samples, a slight excess of yttrium is observed, with a Y/Fe ratio of approximately 0.8 (Figs. \ref{fig:eds_GGG} and \ref{fig:eds_SGGG}). This deviation is consistent with the observed volume expansion and has not been previously reported—despite the fact that many studies have shown similar XRD patterns \cite{Onbasli2014,Manuilov2009a,Liu2014,CaoVan2018}.

\subsection{Magnetic anisotropies and spin relaxation}

\begin{figure}[ht]
\centering
\subfloat[\centering]{\includegraphics[width=7cm, keepaspectratio]{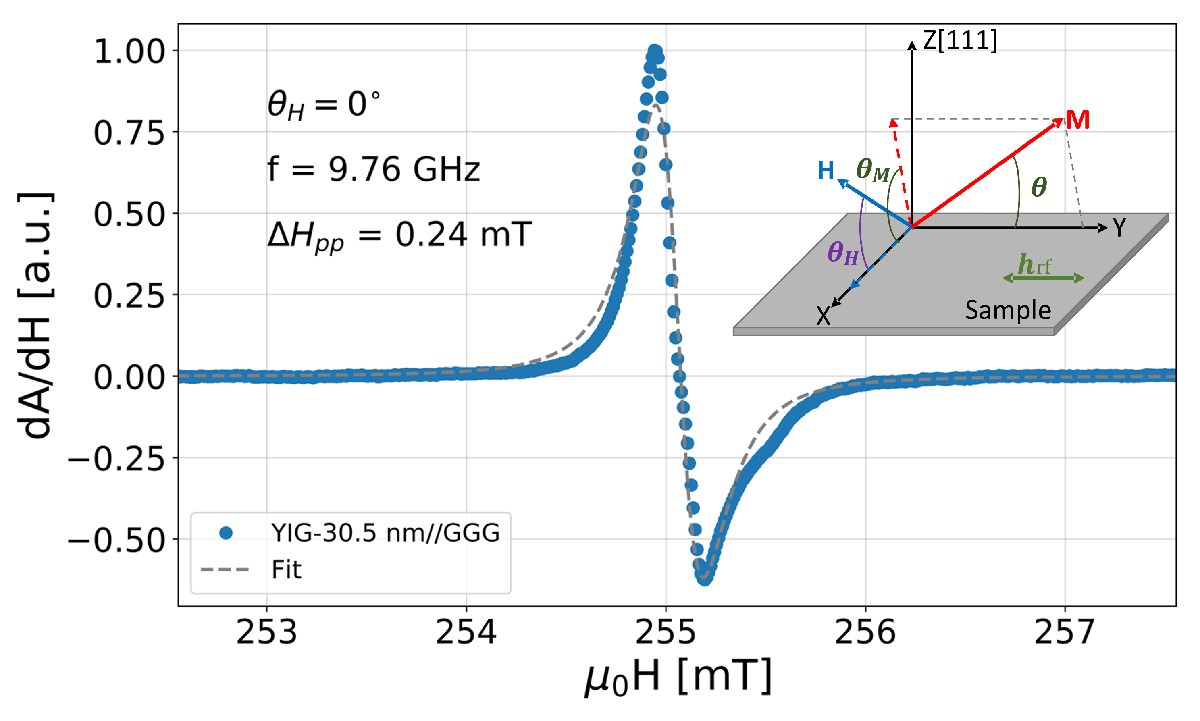}\label{fmr0}}
\subfloat[\centering]{\includegraphics[width=7cm, keepaspectratio]{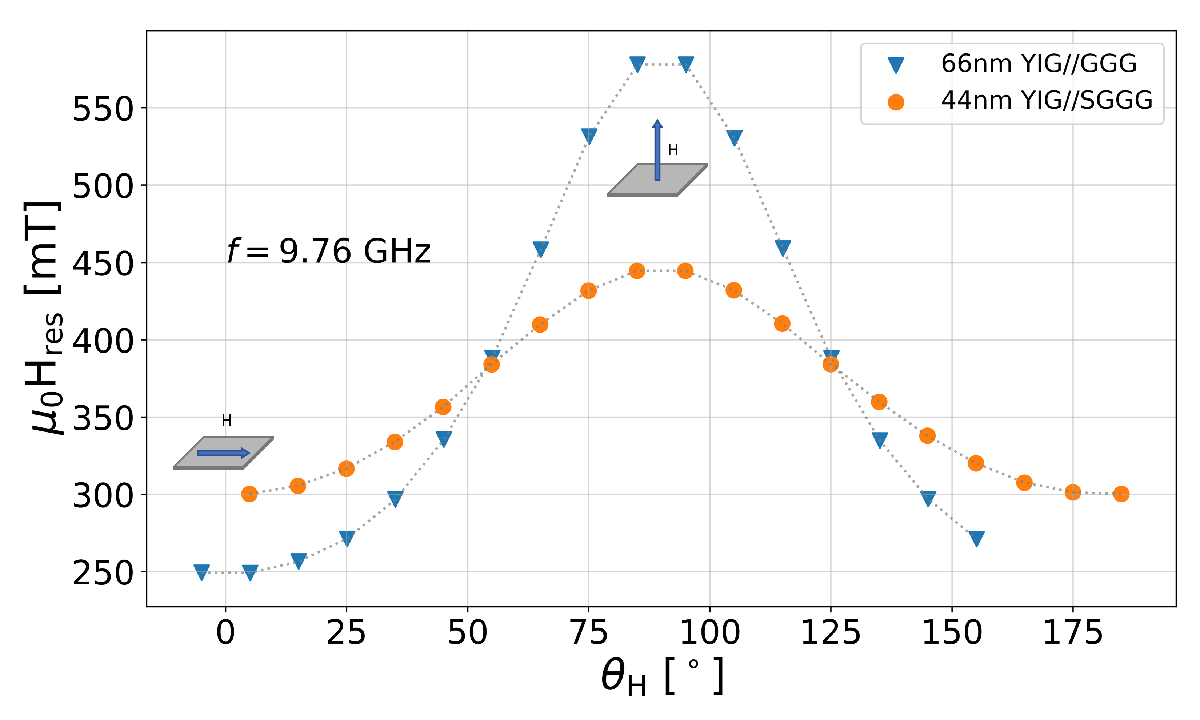}\label{fmr2}}
\caption{(a) FMR spectrum of a 30.5-nm YIG//GGG film, measured at $f$= \unit[9.76]{GHz} and in-plane magnetic field. (b) Out-of-plane angular dependence of the resonance field, $\mu_0$$\mathrm{H}_{\mathrm{res}}$, for YIG films grown on GGG and SGGG substrates.}
\end{figure}

Fig. \ref{fmr0} shows an FMR spectrum of a \unit[30.5]{nm} YIG//GGG film, measured at \unit[9.76]{GHz} with the magnetic field applied in the plane of the film ($\theta_\mathrm{H}=0^{\circ}$, see inset). The first derivative of an asymmetric Lorentzian function was applied to each resonance curve to extract the resonance field ($\mu_0$H$_{\mathrm{res}}$) and the full width at half maximum linewidth ($\mu_0$$\Delta\mathrm{H}_{\mathrm{FWHM}}$). The peak-to-peak linewidth ($\mu_0$$\Delta\mathrm{H}_{\mathrm{pp}}$) of the resonance shown in Fig. \ref{fmr0} is as low as \unit[0.24]{mT}, which is consistent with values reported for high-quality YIG films in the literature \cite{Onbasli2014,Schmidt2020,Manuilov2009a,Sun2012}.

The angular dependence of the resonance fields for the 66-nm-YIG//GGG and 44-nm-YIG//SGGG films, measured as the magnetic field is rotated from $\theta_\mathrm{H}=0^{\circ}$ to $\theta_\mathrm{H}=180^{\circ}$, is shown in Fig. \ref{fmr2}. The curves indicate that easy-plane anisotropy dominates in both cases; however, the anisotropy is noticeably stronger in films grown on GGG substrates compared to those deposited on SGGG. The free energy density, F, used to analyze these data, is given by:
\begin{equation}
 \mathrm{F}=-\mu_0\mathrm{M} \mathrm{H}\sin{\theta_\mathrm{H}}\cos{\left(\theta_\mathrm{H}-\theta_\mathrm{M}\right)}+ (\frac{\mu_0\mathrm{M}^2}{2}-\mathrm{K}_{\perp})\sin^2{\theta}\sin^2{\theta_\mathrm{M}}.
\label{free_energy}
\end{equation}
\noindent Here, $\theta$, $\theta_\mathrm{M}$ and $\theta_\mathrm{H}$ correspond to the angles of the magnetization $\vec{\mathrm{M}}$ and the external magnetic field $\vec{\mathrm{H}}$, as illustrated in the inset of Fig. \ref{fmr0}. K$_{\perp}$ is the anisotropy constant of the system, accounting for the effects of substrate-induced strain. The Smit–Beljers equations \cite{Smit1955} were used to numerically determine the magnetic field dependence of the resonant frequency. For the free energy given in Eq. \ref{free_energy} and under the geometrical configuration described in the Experimental Methods section, the resonance condition takes the following form:
\begin{equation}
\left(\frac{2\pi f}{\gamma}\right)^2=\left[\mu_0\mathrm{H}\cos{\left(\theta_\mathrm{M}-\theta_\mathrm{H}\right)}-\mu_0\mathrm{H}_{\mathrm{eff}}\sin^2{\theta_\mathrm{M}}\right]\times\left[\mu_0\mathrm{H}\cos{\left(\theta_\mathrm{M}-\theta_\mathrm{H}\right)}+\mu_0\mathrm{H}_{\mathrm{eff}}\cos{\left(2\theta_\mathrm{M}\right)}\right].
\label{ecuacionx}
\end{equation}
\noindent where $\mathrm{H}_{\mathrm{eff}}$ is the effective anisotropy field defined as $\mathrm{H}_{\mathrm{eff}}=\mathrm{M}_\mathrm{s}-\frac{2K_{\perp}}{\mu_0\mathrm{M}_\mathrm{s}}$. These parameters, $\mathrm{H}_{\mathrm{eff}}$ and the gyromagnetic ratio $\gamma$, were derived from the fit of the angular dependence of the resonance field using Eqs. \ref{free_energy} and \ref{ecuacionx} (see Fig. \ref{fmr2}, dotted lines). The calculated factor $g=\frac{\gamma \hbar}{\mu_\mathrm{B}}=2.01$ was found to be similar to previously reported values and independent of the thickness of the films and the substrate \cite{Chen2019}.  Due to the strong paramagnetic signal of the Gd ions in the substrates, which accounts for approximately \unit[99.96]{\%} of the signal measured in conventional VSM measurements, the bulk saturation magnetization M$_\mathrm{s}$ was used for calculations. The strain-induced anisotropy field was determined as H$_\mathrm{K} = \frac{2\mathrm{K}_{\perp}}{\mu_0\mathrm{M}_\mathrm{s}}$, along with the anisotropy constant K$_{\perp}$. The effective field and the anisotropy constant for the samples are summarized in Table \ref{datostabla}. 

\begin{table}
\caption{\label{datostabla} Structural and magnetic properties of YIG films grown on GGG and SGGG substrates.}
\begin{tabular}{ccccc}
\hline
Substrate&t$_{\mathrm{YIG}}$ [nm]&$\epsilon_{\perp}$ [\%]&$\mu_0$H$_{\mathrm{eff}}$ [mT]&K$_{\perp}$ ($\times$ 10$^3$) $\left[\frac{\mathrm{J}}{\mathrm{m}^3}\right]$\\
\hline
GGG & 16 & 0.87& 227$\pm$1 &  -3.6$\pm$0.1 \\
GGG & 30.5 & 2.03 & 240$\pm$1 & -4.5$\pm$0.1 \\
GGG & 66 & 1.59 & 239$\pm$1 &  -4.5$\pm$0.1\\
SGGG & 44 & -0.07 & 102$\pm$7 &  5.1 $\pm$0.5 \\
SGGG & 106.5 & -0.34& 96$\pm$5 &  5.5$\pm$0.3 \\
\hline
\end{tabular}
\end{table}

A negative K$_{\perp}$ was obtained for films grown on GGG, indicating easy-plane anisotropy, while a positive K$_{\perp}$ was found for films deposited on SGGG, suggesting the presence of perpendicular magnetic anisotropy. These results correlate with the change in the sign of the lattice mismatch between the films grown on GGG and SGGG substrates.

The anisotropy observed in K$_{\perp}$ is attributed to a strain-induced magnetoelastic anisotropy term (K$_\mathrm{me}$), which can be estimated as \cite{Hansen1984}:
\begin{equation}
\mathrm{K}_{\mathrm{me}}=\frac{3}{2}\lambda_{111}\frac{\mathrm{E}}{2\nu}\epsilon_\perp,
\end{equation}
\noindent where $\lambda_{111}=-2.4\times10^{-6}$ is the magnetoelastic coefficient, and E$=$\unit[$2\times10^{11}$]{Pa} and $\nu=0.29$ are the Young’s modulus and Poisson’s ratio, respectively \cite{Fu2017,Gibbons1958}.

Since strain and stress are not uniform across the film, as evidenced by the GPA images, this calculation provides an upper limit for the magnetoelastic anisotropy: $K_{\mathrm{me}}=\unit[-1.1\times10^3]{\frac{\mathrm{J}}{\mathrm{m}^3}}$ for GGG and $K_{\mathrm{me}}=\unit[13\times10^3]{\frac{\mathrm{J}}{\mathrm{m}^3}}$ for SGGG. This upper limit is consistent with the experimentally obtained values and offers a reasonable estimate of the strain-induced contribution to the out-of-plane anisotropy observed in the SGGG-based films.

\begin{figure}[htbp]
\centering
\includegraphics[width=0.5\textwidth, keepaspectratio]{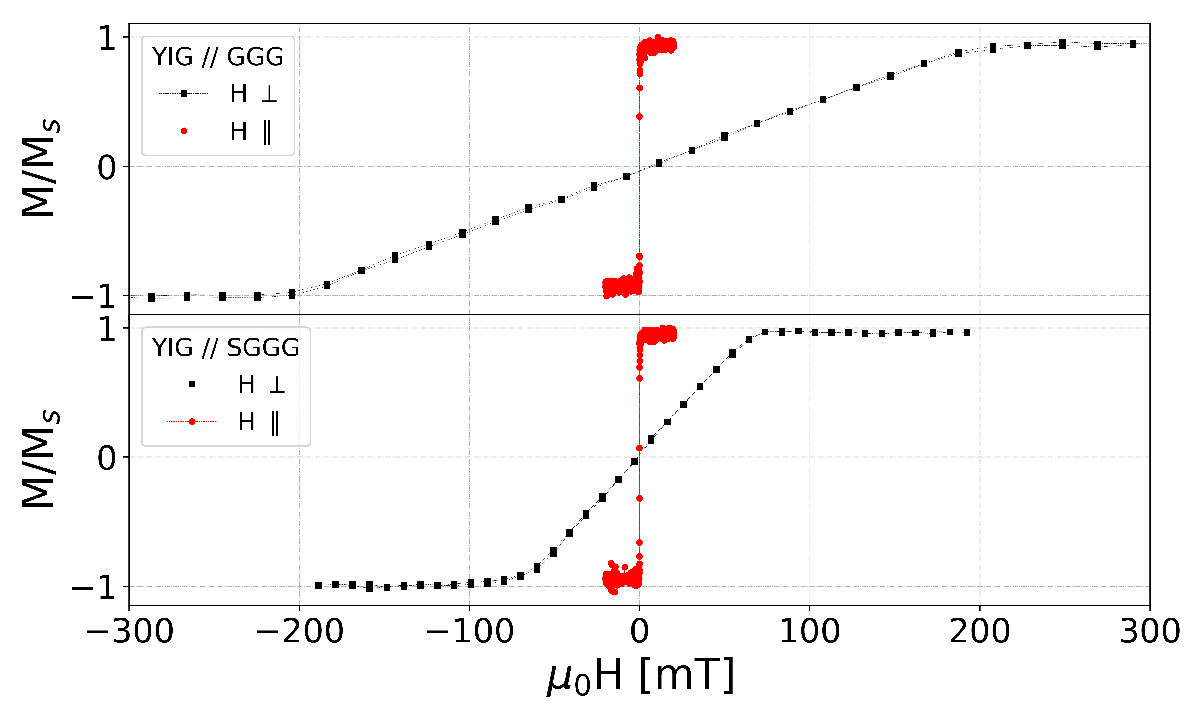}
\caption{OOP and IP magnetization loops for YIG//GGG films (top) and YIG//SGGG films (bottom)}
\label{moke}
\end{figure}

Fig. \ref{moke} shows the out-of-plane (OOP) and in-plane (IP) magnetization loops of YIG films grown on GGG and SGGG substrates measured by MOKE. Both sets of films exhibit an effective easy-plane anisotropy, as indicated by the comparison of OOP and IP curves. However, in agreement with the FMR results, a notable decrease in the saturating field observed in the OOP loops for samples grown on SGGG substrates suggests the presence of a significant perpendicular magnetic anisotropy term in these samples. In the sample grown on the SGGG substrate, the OOP saturating field is reduced to \unit[70]{mT}, while for films grown on the GGG substrate it is \unit[190]{mT}. These values are consistent with the effective magnetic field ($\mu_0$H$_{\mathrm{eff}}$) extracted from the FMR measurements and align well with previously reported data \cite{Ding2020}.

\begin{figure}[htbp]
\centering
\includegraphics[width=0.5\textwidth, keepaspectratio]{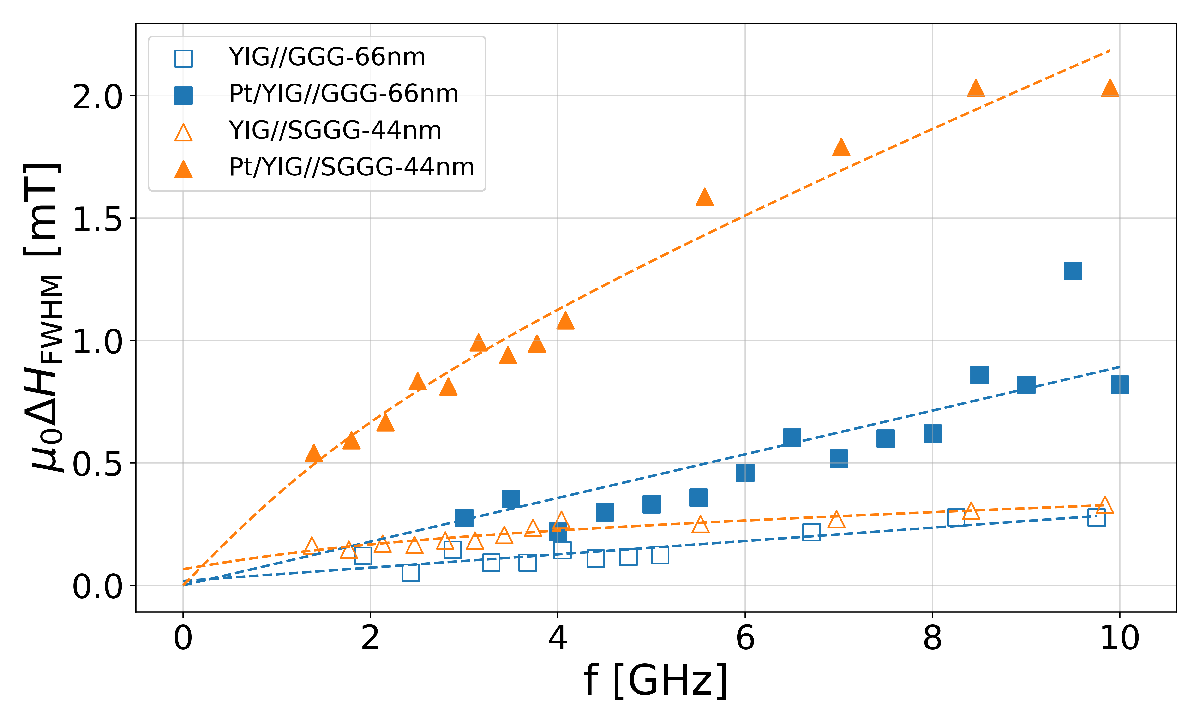}
\caption{Frequency dependence of the ferromagnetic resonance linewidth for \unit[66]{nm}-YIG//GGG and \unit[8]{nm}-Pt/\unit[44]{nm}-YIG//SGGG samples.}
\label{fmr3}
\end{figure}

We also investigated the effect of the substrate on the resonance absorption linewidth. The impact of Pt capping on the FMR linewidth of the \unit[106.5]{nm}-YIG//SGGG film is shown in Fig. \ref{fmr3}. A linear dependence of $\Delta H_{\mathrm{FWHM}}$ on frequency ($f$) was observed in measurements performed on YIG films both with and without Pt capping. However, slight deviations from linearity, particularly in the YIG films grown on SGGG substrates, suggest additional extrinsic contributions such as two-magnon scattering. In such cases, the total linewidth can be expressed as:
\begin{equation}
\Delta\mathrm{H}_{\mathrm{FWHM}} = \frac{2\pi \alpha}{\gamma} f + \Delta H_0 + \Delta H_{2\mathrm{mag}}(f).
\label{gilbert}
\end{equation}
Here $\alpha$ is the Gilbert damping constant, $\Delta \mathrm{H}_0$ represents the inhomogeneous broadening of the linewidth and $\Delta H_{2\mathrm{mag}}(f)$ accounts for frequency-dependent magnon–magnon interactions \cite{Hurben1998}. The Gilbert damping constants extracted from the measurements are listed in Table \ref{datostabla2}. These values are consistent with those previously reported for YIG thin films \cite{Onbasli2014}.

It is well known that platinum deposition on YIG introduces an additional relaxation mechanism \cite{Rezende2013}, whereby angular momentum is transferred from the magnetic film into the adjacent non-magnetic (NM) layer. This spin pumping effect enhances the damping constant by an amount $\Delta\alpha$, which is given by \cite{Tokac2015}:
\begin{equation}
 \Delta\alpha= \frac{\hbar\gamma}{\mathrm{M}_\mathrm{s}} \frac{g^{\uparrow\downarrow}}{\mathrm{t}_{\mathrm{YIG}}},
 \label{gupdown}
\end{equation}
\noindent In this expression, $\hbar$ is the reduced Planck constant, $g^{\uparrow\downarrow}$ is the spin-mixing conductance and t$_{\mathrm{YIG}}$ is the thickness of the ferrimagnetic layer. The enhancement of the Gilbert damping constant, $\Delta\alpha$, was determined by comparing YIG films with and without the Pt capping layer. Using Eq. \ref{gupdown}, $g^{\uparrow\downarrow}$ was calculated (Table \ref{datostabla2}). The obtained values are consistent with those reported in the literature. To contextualize our results we discuss the effective spin-mixing conductance $g^{\uparrow\downarrow}$ derived from damping enhancement with previously reported data. Our $g^{\uparrow\downarrow}$ span from \unit[1.4$\times10^{18}$]{m$^{-2}$} to \unit[5.6$\times10^{19}$]{m$^{-2}$}. i. e. Qiu et al. found \unit[1.3$\times10^{18}$]{m$^{-2}$} for a well-controlled interface \cite{Qiu2013}, Lustikova et al. reported \unit[2$\times10^{18}$]{m$^{-2}$} for sputtered, post-annealed YIG films \cite{Lustikova2014} and Jungfleisch et al. obtained values on the order of \unit[3.43$\times10^{19}$]{m$^{-2}$} depending on the assumed Pt spin-diffusion length and analysis approach \cite{Jungfleisch2013}, while Haertinger et al. reported $g^{\uparrow\downarrow}$ in the \unit[10$^{18}$]{m$^{-2}$} range with thickness-dependent variations \cite{Haertinger2015}. These values are within the typical range found in Pt/YIG systems (\unit[10$^{18}$-10$^{19}$]{m$^{-2}$}). Comparing our results to these benchmarks, the thin YIG//GGG films (16 and \unit[30.5]{nm}) show $g^{\uparrow\downarrow}$ within the lower-to-mid range of reported values, consistent with Qiu and Lustikova for ex-situ or non-perfect interfaces \cite{Qiu2013,Lustikova2014}. Intermediate-thickness samples (\unit[66]{nm} on GGG and \unit[44]{nm} on SGGG) exhibit substantially larger $g^{\uparrow\downarrow}$, near the upper limit of reported data found in carefully prepared Pt/YIG interfaces \cite{Jungfleisch2015,Haertinger2015}. The thickest sample (\unit[106.5]{nm} on SGGG) reaches the highest $g^{\uparrow\downarrow}$ measured. Several factors contribute to this spread: (i) interface quality, including atomic sharpness and chemical contamination, which can change $g^{\uparrow\downarrow}$ by orders of magnitude \cite{Jungfleisch2013}; (ii) substrate and strain state, which modify magnon dispersion and magnon–phonon coupling and thus affect the reference damping used to extract $g^{\uparrow\downarrow}$ \cite{Rckriegel2014,Schlitz2022,Serha2025,Du2014a} and (iii) non-spin-pumping linewidth contributions, such as two-magnon scattering or inhomogeneity, which can bias $\Delta \alpha$ \cite{Rezende2013,Kapelrud2013}. To further investigate the influence of strain on spin-charge conversion efficiency, ISHE and SSE measurements were performed on these samples, as discussed in the following sections.

\subsection{Inverse Spin Hall and Spin Seebeck Effects}

The substrate-induced strains discussed above correlate with notable changes in both the effective anisotropy fields and damping mechanisms observed in FMR measurements, which in turn influence spin-charge conversion phenomena such as the ISHE \cite{Saitoh2006} and SSE \cite{Uchida2012,Uchida2010}.

\begin{figure}[ht]
\centering
\subfloat[\centering]{\includegraphics[width=7cm, keepaspectratio]{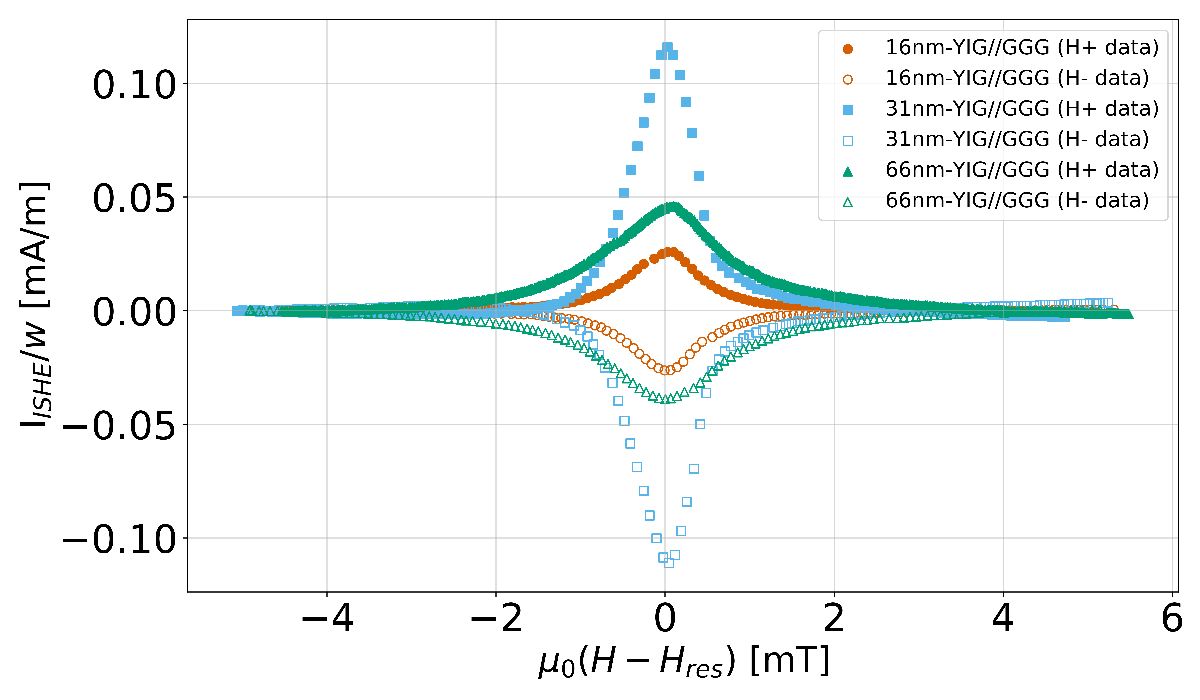}}
\subfloat[\centering]{\includegraphics[width=7cm, keepaspectratio]{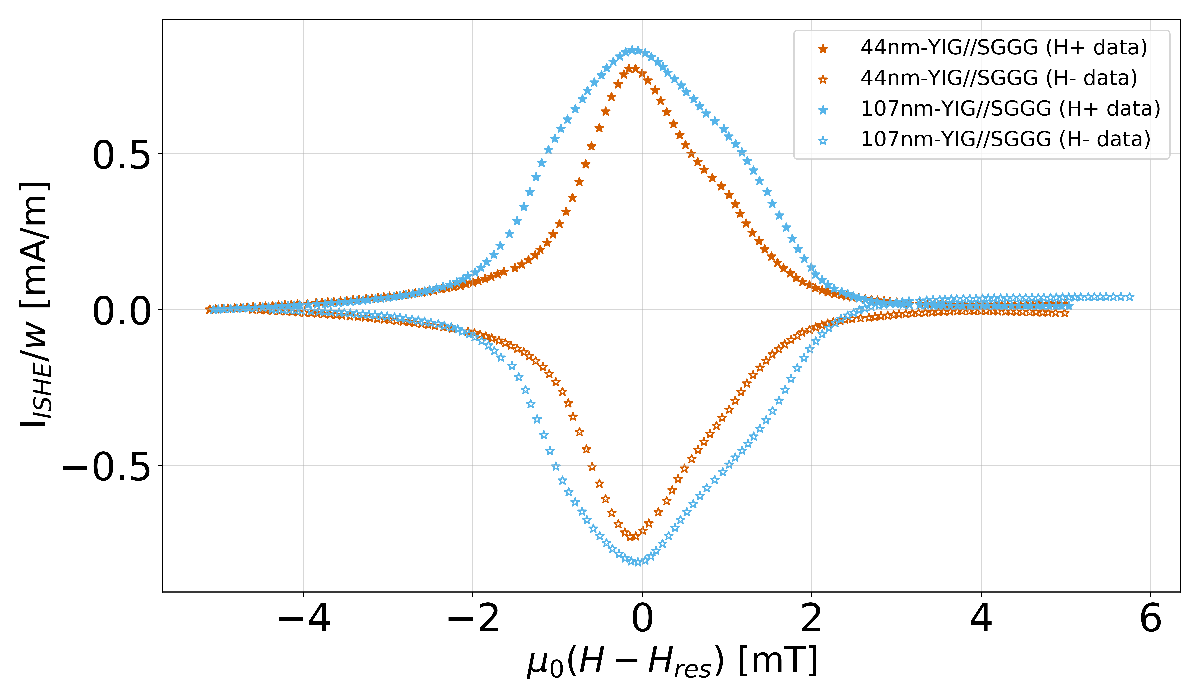}}
\caption{ISHE voltage normalized by resistance and width vs external magnetic field measured on a Pt/YIG sample on (a) GGG and (b) SGGG substrates. The FMR resonance frequency is \unit[9.76]{GHz}.}
\label{ishe_all}
\end{figure}

Fig. \ref{ishe_all} shows a representative ISHE voltage signal for a Pt/YIG bilayer, with the peak fitted using a symmetric Lorentzian function to extract the ISHE contribution. The peak amplitude, denoted V$_{\mathrm{ISHE}}$, corresponds to the resonant magnetic field $\mu_0$H$_{\mathrm{res}}$. The measured I$_{\mathrm{ISHE}}$ values, normalized by the sample width to account for geometric differences, are summarized in Table \ref{datostabla2} for in-plane field orientation ($\theta_\mathrm{H}=0^{\circ}$). The deviation from a single Lorentzian shape and the presence of multiple resonance features in the thicker YIG//SGGG samples are primarily attributed to magnetic inhomogeneities and two-magnon scattering processes. These inhomogeneities, along with enhanced two-magnon scattering, broaden and distort the resonance line shape, resulting in the observed multiple-peak ISHE signals \cite{Hauser2016,Ding2020,Haertinger2015}.

\begin{figure}[ht]
%\isPreprints{}{% This command is only used for ``preprints''.
%\begin{adjustwidth}{-\extralength}{0cm}
\centering
%} % If the paper is ``preprints'', please uncomment this parenthesis.
\subfloat[\centering]{\includegraphics[width=7cm, keepaspectratio]{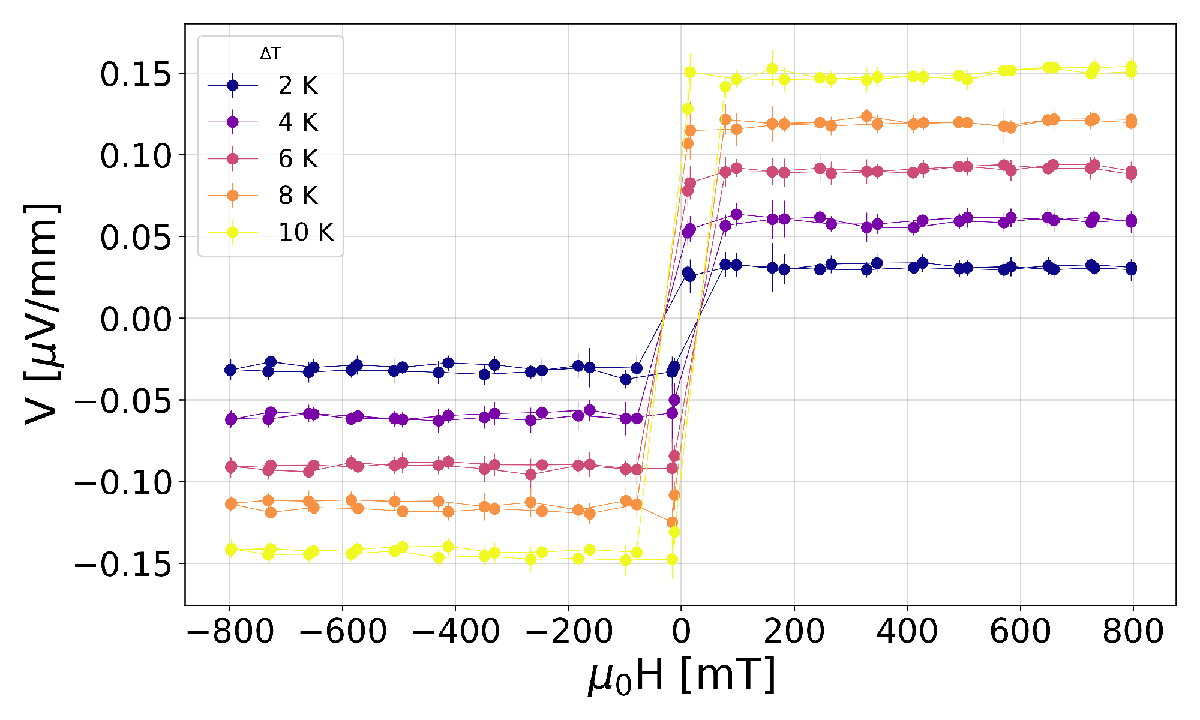}\label{sse1}}
\subfloat[\centering]{\includegraphics[width=7cm, keepaspectratio]{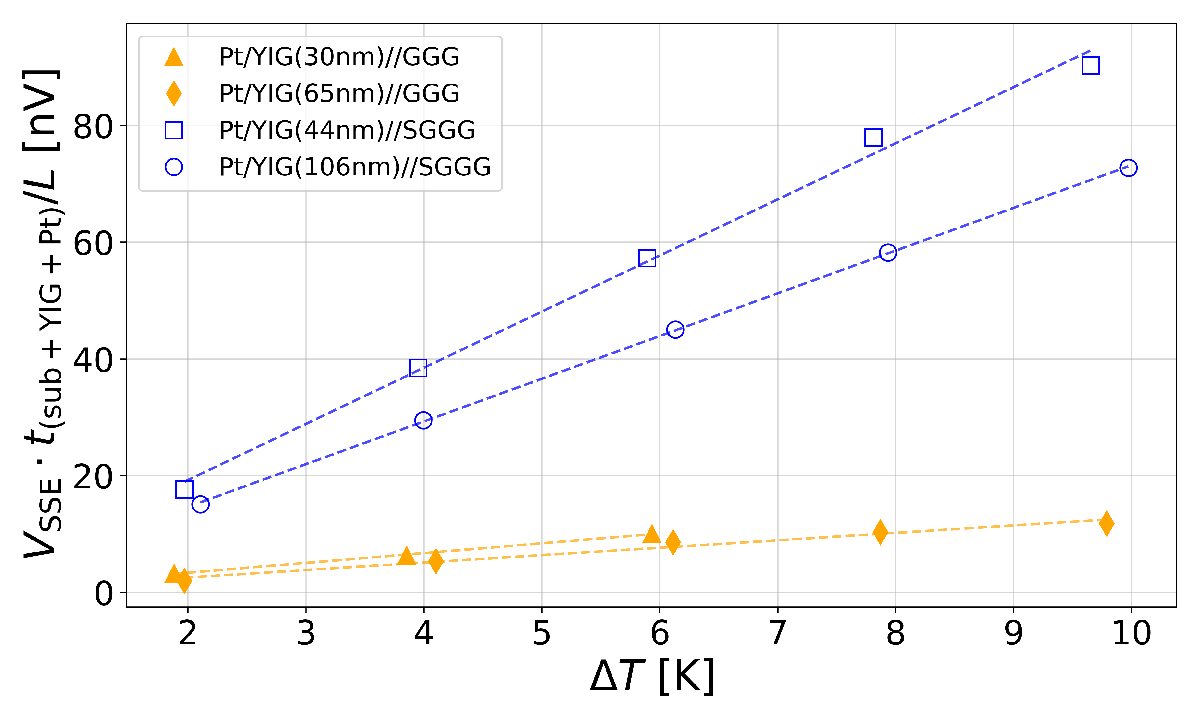}\label{sse2}}
%\isPreprints{}{% This command is only used for ``preprints''.
%\end{adjustwidth}
%} % If the paper is ``preprints'', please uncomment this parenthesis.
\caption{(a) Spin Seebeck Effect voltage normalized by length vs magnetic field and temperature gradient perpendicular to the surface measured on the Pt/YIG samples on different substrates at room temperature. (b) Voltage generated by SSE, normalized by the substrate plus film thickness and sample length as a function of the applied temperature gradient.}
\end{figure}

YIG films grown on GGG substrates, subjected to tensile out-of-plane strains, exhibited ISHE voltages nearly an order of magnitude lower than those on SGGG substrates, which are subject to slight compressive strain. Similarly, SSE measurements reveal a transverse voltage in the Pt layer, arising from the conversion of thermally generated spin currents via the ISHE. Fig. \ref{sse1} shows a representative voltage signal for a Pt/\unit[106.5]{nm}-YIG//SGGG sample, with the curve exhibiting odd symmetry with respect to the magnetic field and a linear dependence on the applied temperature gradient (Fig. \ref{sse2}). 

To quantify the coupling between spin and heat transport, the SSE coefficient is defined as:
\begin{equation}
\mathrm{S}_{\mathrm{SSE}} = \frac{\mathrm{V}_{\mathrm{SSE}}}{\mathrm{L}} \frac{\mathrm{t}_{\mathrm{Pt+YIG+sub}}}{\Delta \mathrm{T}},
\label{SSEcoefficient}
\end{equation}
\noindent where V$_{\mathrm{SSE}}$ is the maximum transverse voltage, $\Delta\mathrm{T}$ is the temperature difference across the sample, and $\mathrm{t}_{\mathrm{YIG+sub}}$ is the total thickness of the YIG film and substrate. The sample length and width are L$=$\unit[7]{mm} and $\mathrm{w}=$\unit[4]{mm}, respectively. The calculated $\mathrm{S}_{\mathrm{SSE}}$ coefficients for all samples are summarized in Table  \ref{datostabla2}.

\begin{table}
\caption{\label{datostabla2} V$_{\mathrm{ISHE}}$ and the S$_{\mathrm{SSE}}$ coefficients measured for Pt/YIG films grown on GGG and SGGG substrates at room temperature}
\begin{tabular}{ccccccc}
\hline
Substrate&t$_{\mathrm{YIG}}$&$\alpha$&$\alpha_{Pt}$&$g^{\uparrow\downarrow}$($\times10^{19}$)&I$_{\mathrm{ISHE}}$/w&S$_{\mathrm{SSE}}$\\
&[nm]&($\times10^{-4}$)&($\times10^{-4}$)&[m$^{-2}$]&[$\mu$A.m$^{-1}$]&[nV.K$^{-1}$]\\
\hline
GGG&16&1.5$\pm$0.9&13.3$\pm$0.9&0.18$\pm$0.03&25.7$\pm$0.3&0\\
GGG&30.5&10$\pm$1&14.8$\pm$0.8&0.14$\pm$0.06&115.7$\pm$0.3&1.7$\pm$0.1\\
GGG&66&3.5$\pm$0.6 &20.4$\pm$0.7&1.05$\pm$0.08&42$\pm$8&1.3$\pm$0.1 \\
SGGG&44&3.7$\pm$0.4&30$\pm$2&1.1$\pm$0.1&750$\pm$60&9.6$\pm$0.4\\
SGGG&106.5&2.7$\pm$0.4&57$\pm$2&5.6$\pm$0.2&820$\pm$50&7.32$\pm$0.05\\
\hline
\end{tabular}
\end{table}

YIG films grown on SGGG substrates consistently show the highest V${\mathrm{ISHE}}$ and S${\mathrm{SSE}}$ values, along with a noticeable perpendicular anisotropy component induced by substrate strain. This demonstrates the crucial role of strain in enhancing spin-charge conversion efficiency. A quantitative analysis of the damping and spin-to-charge conversion signals reveals that damping alone cannot explain the observed differences in ISHE and SSE signals between samples grown on GGG and SGGG substrates. Despite similar Gilbert damping values, samples on SGGG exhibit ISHE and SSE voltages up to an order of magnitude higher. The enhanced $g^{\uparrow\downarrow}$ values in these samples indicate improved spin transparency at the Pt/YIG interface, likely arising from a combination of strain-induced anisotropy and high-quality interfacial structure. STEM-HAADF analysis confirms well-defined Pt/YIG interfaces in both systems, suggesting that strain may facilitate more favorable interfacial bonding or reduce spin scattering. These findings are consistent with prior reports \cite{Liu2020} and highlight the importance of interface engineering and strain control in optimizing spin-charge conversion efficiency for spin-orbitronic applications.

%%%%%%%%%%%%%%%%%%%%%%%%%%%%%%%%%%%%%%%%%%
\section{Conclusions}

In this work, we presented a comprehensive study of spin pumping in Pt/YIG bilayers, with a focus on the influence of substrate-induced strain. By comparing YIG films grown on GGG and SGGG substrates, we investigated the effects of tensile and unexpected compressive out-of-plane strains on spin transport and spin-to-charge conversion. This highlights the complex interplay between structural distortion and magnetic properties in epitaxial YIG films.

Structural and compositional analyses indicated deviations from volume conservation in YIG//GGG films, characterized by tensile out-of-plane strain and slight Y enrichment, despite minimal lattice mismatch. These results suggests that local defects, such as Y antisite substitutions, contribute to strain. These microstructural features correlate with the observed variations in magnetic damping and anisotropy.

Magnetic measurements showed that substrate-induced strain modifies anisotropy fields, promoting in-plane anisotropy in GGG-grown films and inducing a perpendicular anisotropy component in SGGG-based samples. Spin pumping experiments, using both microwave and thermally generated spin currents, demonstrated a substantial enhancement in spin-to-charge conversion efficiency for SGGG-grown YIG. This improvement is attributed to strain-mediated increases in the spin mixing conductance $g^{\uparrow\downarrow}$ and enhanced spin transparency at the Pt/YIG interface.

Overall, these results put in evidence that controlled substrate-induced strain alongside high-quality interface engineering provide an effective route for tuning spin transport and optimizing spin-charge conversion efficiency in YIG-based spintronic devices. This approach offers a promising framework for designing strain-engineered magnetic insulators with tailored spin–orbit coupling performance.

\ack{This work was partially supported by PICT-2019-02781 ANPCyT and PICT-0415 ANPCyT, Argentina, H2020-MSCA-RISE-2016 Project n$^{\circ}$ 734187 SPICOLOST and MSCA-RISE-2021 Project n$^{\circ}$ 101007825 ULTIMATE-I. Carlos Garc\'ia acknowledges financial support from ANID FONDECYT/REGULAR 1201102, ANID FONDECYT/REGULAR 1241918 and ANID FONDEQUIP EQM140161. Mar\'ia Abell\'an acknowledge the financial support received by ANID PIA/APOYO AFB230003. The authors would like to thank the ``Laboratorio de Microscop\'ias Avanzadas" and the ``Servicio General de Apoyo a la Investigaci\'on" at ``Universidad de Zaragoza" for providing access to their instruments. We also thank Claudio Bonin from the ``Instituto de F\'isica del Litoral (CONICET-UNL)" for performing the MOKE measurements. F. B. acknowledge financial support of the European Union under the REFRESH – Research Excellence For REgion Sustainability and High-tech Industries project n$^{\circ}$ CZ.10.03.01/00/22/003/0000048 via the Operational Programme Just Transition and ENREGAT supported by M\v{S}MT, project n$^{\circ}$ LM2023056.}

\bibliographystyle{unsrt}
\bibliography{iopjournal-template}

\end{document}